\documentclass[12pt]{iopart}

\usepackage[utf8]{inputenc}
\usepackage[T1]{fontenc}
\usepackage{iopams}
\usepackage{amsfonts}
\usepackage{amssymb}
\usepackage{graphicx}
\usepackage{hyperref}
\hypersetup{
	colorlinks=true,
	hidelinks,
	linkcolor=blue,
	filecolor=magenta,      
	urlcolor=cyan,
	pdftitle={Complete Document for Review (PDF Only)},
	pdfpagemode=FullScreen,
}
\usepackage{mleftright}
\usepackage{placeins}

\begin{document}
\title[Optimal sensor fusion method for active vibration isolation systems]{Optimal sensor fusion method for active vibration isolation systems in ground-based gravitational-wave detectors}
\author{
	T.T.L. Tsang$^{1, a}$,
	T.G.F. Li$^{1, 2, 3}$,
	T.Dehaeze$^{4}$,
	C. Collette$^{4}$
}
\address{${}^{1}$Department of Physics, The Chinese University of Hong Kong, Shatin, New Territories, Hong Kong}
\address{${}^{2}$Institute for Theoretical Physics, KU Leuven, Celestijnenlaan 200D, B-3001 Leuven, Belgium}
\address{${}^{3}$Department of Electrical Engineering (ESAT), KU Leuven, Kasteelpark Arenberg 10, B-3001 Leuven, Belgium}
\address{${}^{4}$Precision Mechatronics Laboratory, University of Liege, Belgium}
\ead{${}^{a}$\href{mailto:ttltsang@link.cuhk.edu.hk}{ttltsang@link.cuhk.edu.hk}}
\vspace{10pt}
	
\begin{abstract}
Sensor fusion is a technique used to combine sensors with different noise characteristics into a super sensor that has superior noise performance.
To achieve sensor fusion, complementary filters are used in current gravitational-wave detectors to combine relative displacement sensors and inertial sensors for active seismic isolation.
Complementary filters are a set of digital filters, which have transfer functions that are summed to unity.
Currently, complementary filters are shaped and tuned manually rather than being optimized.
They can be sub-optimal and hard to reproduce for future detectors.
In this paper, $\mathcal{H}_\infty$ optimization is proposed for synthesizing optimal complementary filters.
The complementary filter design problem is converted into an optimization problem that seeks minimization of an objective function equivalent to the maximum difference between the super sensor noise and the lower bound in logarithmic scale.
The method is exemplified with three cases, which simulate the sensor fusion between a relative displacement sensor and an inertial sensor.
In all cases, the $\mathcal{H}_\infty$ complementary filters suppress the super sensor noise equally close to the lower bound at all frequencies in logarithmic scale.
The $\mathcal{H}_\infty$ filters also provide better suppression of sensor noises compared to complementary filters pre-designed using traditional methods.
\end{abstract}

%
\vspace{2pc}

\submitto{\CQG}

 
%
	
	\section{Introduction}
Vibration noise, such as seismic noise, is one of the major noise sources for ground-based large-scale scientific instruments like interferometric gravitational wave detectors.
Current gravitational-wave detectors, such as interferometric gravitational-wave detector like LIGO~\cite{aligo}, Virgo~\cite{Acernese_2014}, and KAGRA~\cite{10.1093/ptep/ptx180, Akutsu_2019}, use multi-stage pendulums to suspend the core optics of the interferometers, passively isolating them from high-frequency external vibration in the detection band (10-1000 Hz)~\cite{quadpaper, update_quadruple, Braccini:2000hug, KAGRA:2019ywe, typeb_paper}.
On top of that, the pendulums are typically mounted on isolation platforms equipped with sensors and actuators to actively isolating the vibration noise at lower frequencies (<10 Hz) and damping the resonances of the pendulums~\cite{Braccini:2000hug,typeb_paper,Matichard:2015eva}.
External seismic disturbance, such as the microseism~\cite{microseism}, at lower frequencies can cause the suspended optics to move excessively.
This will cause the instruments to misalign, and in severe cases, result in the temporary shutdown of the detectors.
In fact, low frequency seismic noise has ultimately limited the duty cycle of the LIGO and KAGRA detectors~\cite{lens.org/141-760-662-241-570, LIGO:2020pzq}.
Therefore, active isolation is extremely important in these large-scale instruments and it remains as an active research topic in the field of experimental physics.

Active isolation comes with the price of control noise addition.
Control noise can be injected to the detector, compromising the sensitivity of the detector, and must be limited.
One way to reduce the control noise is to lower the noises of the sensors used for feedback control in active isolation.
Recent research has been made to develop low-noise inertial sensors for active isolation systems in gravitational-wave detectors \cite{inclinometer, compact_interferometer,2019_6d_interferometric_inertial_isolation_system}.
Inertial sensors can be used to achieve active isolation.
They have good noise performance at higher frequencies but have poor performance at lower frequencies and could cause isolation platforms to drift.
On the other hand, relative displacement sensors, such as linear variable differential transformers (LVDTs) used in Ref.~\cite{typeb_paper}, have lower noise at low frequencies.
But they only measure relative displacement so they cannot be used for active isolation.
However, it is possible to utilize both sensors together via a control strategy called sensor fusion.
This way, active seismic noise isolation can be achieved using the inertial sensors without drift.

Sensor fusion is a technique that combines two or more sensors to form a so-called ``super sensor'' that can have superior noise characteristics than the individual sensors.
In this case, we assume the sensors to measure a common signal but have individual uncorrelated intrinsic sensor noises.
There are multiple ways to achieve sensor fusion, such as the use of Kalman filters~\cite{robert12_introd_random_signal_applied_kalman, brown72_integ_navig_system_kalman_filter} and complementary filters~\cite{plummer06_optim_compl_filter_their_applic_motion_measur}.
In particular, complementary filters have been widely used in active isolation platforms in gravitational wave detectors~\cite{Matichard:2015eva,LIGO:2020pzq,hua05_low_ligo,Hua04polyphasefir,ligo_earthquake_mode, Sekiguchi:2016cdw, Heijningen2018TurnUT,lucia_thesis,fujii_thesis}.
Complementary filters are a set of digital filters that can take almost any arbitrary shape so long as their transfer functions are summed to unity.
The design of their shapes is important as it determines the final noise performance of the super sensor.
Although the technique is used in current gravitational-wave detectors, the design methodology was not thoroughly discussed and the filters designed were either sub-optimal and were hardly reproducible.
While the method will be adopted by detectors like KAGRA~\cite{typeb_paper} and the Einstein Telescope~\cite{ET_design_report_update_2020}, it would be convenient to have a method to optimize complementary filters that minimize the noise floor of the super sensors.

Past research has used particle swarm optimization to design sensor correction filters at LIGO \cite{pso_sensor_correction}.
And, it was already shown that the detector can benefit from improving control filter designs.
While numerical optimization approaches can be used to optimize complementary filters and other control filters, there are certain limitations.
The numerical approach requires control filters to take a specified form of transfer function, e.g. 4 pairs of complex poles and 3 pairs of complex zeros in Ref.~\cite{pso_sensor_correction}.
This effectively limits the full parameter space to a much smaller subspace for optimization.
This is because control filters can have almost any arbitrary number of simple and complex poles and zeros as long as it is stable and proper.
This would mean that filters from such an optimization approach may not truly be optimal, if the optimal filter does not fall into the subspace that the numerical optimization takes place.

Recently, a complementary filter shaping method using $\mathcal{H}_\infty$ synthesis was proposed~\cite{dehaeze19_compl_filter_shapin_using_synth}.
In the work, it was shown that $\mathcal{H}_\infty$ methods can be used to optimize complementary filters that satisfy frequency-dependent filter shape specifications.
In contrast to numerical optimization approaches, $\mathcal{H}_\infty$ optimization assumes no predefined structure of the filters.
In this paper, we will extend the idea and propose to use $\mathcal{H}_\infty$ methods to synthesize optimal complementary filters in such a way the super sensor noise is minimized.

This paper is structured as follows.
Sec.~\ref{sec:method} gives an overview to the sensor fusion technique using complementary filters and introduces $\mathcal{H}_\infty$ methods for complementary filter optimization.
In Sec.~\ref{sec:result}, the optimization of complementary filters for the fusion of typical sensors used in active isolation platforms of gravitational-wave detectors is demonstrated.
In Sec.~\ref{sec:discussion}, some discussions regarding the proposed method, limitations, and future works are noted.
In Sec.~\ref{sec:conclusion}, a summary of this paper is given.

	\section{Methodology \label{sec:method}}
\subsection{Sensor fusion using complementary filters}
Without loss of generality, let's define complementary filters to be a set of filters whose transfer functions are summed to unity, i.e.
\begin{equation}
	\sum_i H_i\mleft(s\mright) = 1\,,
	\label{eqn:complementary_filter_unity}
\end{equation}
where $H_i$ is the transfer function of the $i^\mathrm{th}$ filter, and $s$ is a complex variable.
Each complementary filter is filtering the output of the individual sensors and the filtered signals are summed.
The super sensor readout $X_\mathrm{super}(s)$ is therefore given by
\begin{equation}
	X_\mathrm{super}\mleft(s\mright) = \sum_i H_i\mleft(s\mright)X_i\mleft(s\mright)\,,
	\label{eqn:super_sensor_readout}
\end{equation}
where $X_i(s)$ is the sensor readout of the $i^\mathrm{th}$ sensor.
Each sensor is modeled as having additive sensing noise $N_i(s)$.
So, each sensor readout $X_i(s)$ reads
\begin{equation}
	X_i\mleft(s\mright) = X\mleft(s\mright) + N_i\mleft(s\mright)\,,
	\label{eqn:sensor_readout}
\end{equation}
where $X(s)$ is the common signal that the sensors are all measuring.
Substituting Eqn.~\ref{eqn:complementary_filter_unity} and Eqn.~\ref{eqn:sensor_readout} into Eqn.~\ref{eqn:super_sensor_readout}, we get
\begin{equation}
	X_\mathrm{super}\mleft(s\mright) = X\mleft(s\mright)+\sum_i H_i\mleft(s\mright)N_i\mleft(s\mright)\,.
	\label{eqn:super_sensor_readout_with_noise}
\end{equation}
The super sensor readout $X_\mathrm{super}(s)$ is then equal to the common signal $X(s)$ plus the noise of each sensor filtered out by the complementary filters.
As the sensors have intrinsic noise $N_i(s)$ with different frequency content, the goal of sensor fusion is to design a set of complementary filters $H_i(s)$ that achieve optimal trade-off between these sensor noises at different frequencies.

\subsection{$\cal{H}_\infty$ synthesis}
\begin{figure}[!h]
	\centering
	\def\svgwidth{0.5\columnwidth}
	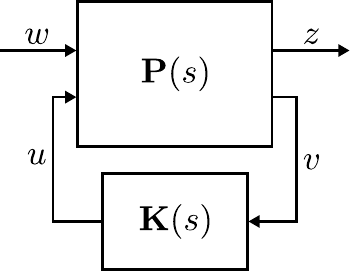
	\caption{Generalized plant representation}
	\label{fig:generalized_plant_representation}
\end{figure}

In $\cal{H}_\infty$ methods~\cite{zames_hinfinity}, a system is specified in the generalized plant representation as shown in Fig.~\ref{fig:generalized_plant_representation}.

The augmented plant $\mathbf{P}$ has two inputs and two outputs.
The inputs $w$ and $u$ are the exogenous inputs and the manipulated variables respectively. 
And, the outputs $z$ and $v$ are the error signals and the measurements respectively.
Note that these variables are vector valued in general.

In the open loop configuration, the input-output relation reads
\begin{equation}
	\left(
	\begin{array}{c}
		z\\
		v
	\end{array}
	\right)
	=
	\mathbf{P}\mleft(s\mright)
	\left(
	\begin{array}{c}
		w\\
		u
	\end{array}
	\right)
	=
	\left[
	\begin{array}{cc}
	P_{11}\mleft(s\mright) & P_{12}\mleft(s\mright)\\
	P_{21}\mleft(s\mright) & P_{22}\mleft(s\mright)\\
	\end{array}
	\right]
	\left(
	\begin{array}{c}
		w\\
		u
	\end{array}
	\right)
	\,.
\end{equation}
When the loop is closed, the manipulated inputs $u$ are generated from the measured output $v$ via a regulator $\mathbf{K}\mleft(s\mright)$,
\begin{equation}
	u=\mathbf{K}\mleft(s\mright)v\,.
\end{equation}
In such configuration, the input-output relation becomes
\begin{equation}
	z=\left[P_{11}(s)+P_{12}(s)\mathbf{K}(s)\left(I-P_{22}(s)\mathbf{K}(s)\right)^{-1}P_{21}(s)\right]w\,,
	\label{eqn:input_output_relation}
\end{equation}
and the closed loop transfer function matrix is defined as
\begin{equation}
	\mathbf{G}\mleft(s;\mathbf{K}(s)\mright)\equiv P_{11}(s)+P_{12}(s)\mathbf{K}(s)\left(I-P_{22}(s)\mathbf{K}(s)\right)^{-1}P_{21}(s)\,,
\end{equation}
where $I$ is the identity matrix.

$\mathcal{H}_\infty$ methods are used to synthesize $\mathcal{H}_\infty$-optimal controllers for feedback systems.
$\mathcal{H}_\infty$-optimal controller are stabilizing regulator that minimizes the $\mathcal{H}_\infty$ norm of the close-loop transfer function $\mathbf{G}(s;\mathbf{K}(s))$.
So, the $H_\infty$ optimal controller can be seen as
\begin{equation}
	\mathbf{K}_\infty(s) = \textrm{argmin}_{\mathbf{K}(s)\in\mathcal{K}}\left\| \mathbf{G}(s;\mathbf{K}(s))\right\|_\infty\,,
\end{equation} 
where $\mathcal{K}$ is the set of all controllers such that the closed-loop transfer function is stable and $\|\cdot\|_\infty$ denotes the $\mathcal{H}_\infty$ norm of a transfer function.
The $\mathcal{H}_\infty$ norm is defined as
\begin{equation}
	\left\| \mathbf{G}(s;\mathbf{K}(s))\right\|_\infty = \sup_{\omega}\bar\sigma\mleft(\mathbf{G}\mleft(j\omega;\mathbf{K}(s)\mright)\mright)\,,
	\label{eqn:h_infinity_norm}
\end{equation}
where $\omega$ is the angular frequency, $j$ is the imaginary number, and $\bar\sigma(\cdot)$ denotes the maximum singular value.

There are a few ways to approach $\mathcal{H}_\infty$-optimal controller, including Riccati-based methods~\cite{Glover1988StatespaceFF} and LMI-based (linear matrix inequality) methods~\cite{PACKARD1992271}.
In this work, the $\mathcal{H}_\infty$ problems are solved using $\mathcal{H}_\infty$ synthesis function \verb|control.hinfsyn()| in the Python Control Systems library \verb|control|~\cite{python_control}.
This function is a Python wrapper for \verb|SLICOT|~\cite{Benner1999} Fortran subroutine \verb|SB10AD|, which is a function that computes $\mathcal{H}_\infty$ optimal controller using a modified version of the Riccati-based method~\cite{fortran77}.
The complementary filter synthesis method is also made available with an open-source Python package called \verb|Kontrol|~\cite{kontrol}.


\subsection{Complementary filter problem as an $\cal{H}_\infty$ synthesis problem \label{sec:comp_hinf}}
\begin{figure}[!h]
	\centering
	\def\svgwidth{0.7\columnwidth}
	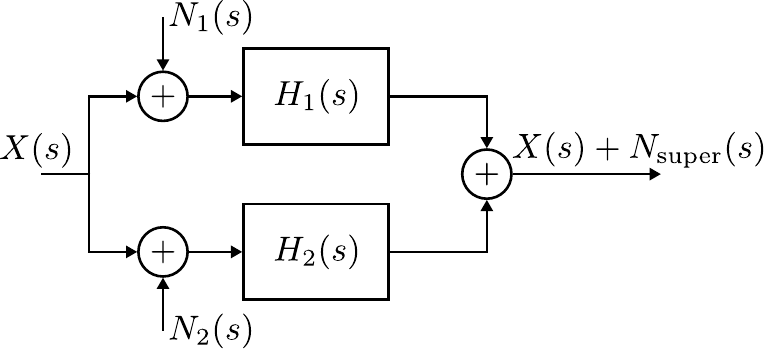
	\caption{Block diagram of a two-sensor complementary filter configuration.}
	\label{fig:two_sensors}
\end{figure}

Consider a two sensors configuration shown in Fig.~\ref{fig:two_sensors}. 
Two sensors are reading the same signal $X(s)$ but with different sensing noises, i.e. $N_1(s)$ and $N_2(s)$.
The two sensor readouts are filtered by complementary filters $H_1\mleft(s\mright)$ and $H_2\mleft(s\mright)$, respectively.
The filtered signals are summed eventually to become a super sensor which has the noise term
\begin{equation}
	N_\mathrm{super}(s) = H_1\mleft(s\mright)N_1(s) + H_2\mleft(s\mright)N_2(s)\,,
	\label{eqn:supersensornoise}
\end{equation}
where $N_\mathrm{super}(s)$ is the sensing noise of the super sensor.

\begin{figure}[!h]
	\centering
	\def\svgwidth{0.7\columnwidth}
	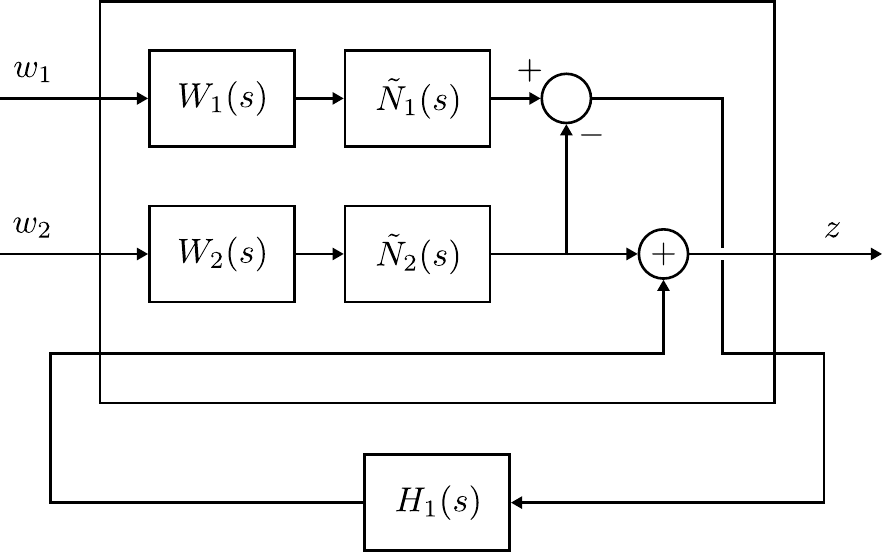
	\caption{Complementary filter configuration augmented as a generalized plant.}
	\label{fig:augmented_plant}
\end{figure}

To convert the complementary filter synthesis problem to an $\mathcal{H}_\infty$ problem, it has to be expressed with the generalized plant representation as shown in Fig.~\ref{fig:generalized_plant_representation}.
The generalized plant representation of a two-sensor complementary filter configuration is shown in Fig.~\ref{fig:augmented_plant}.

The exogenous inputs $w=\left(w_1, w_2\right)^T$ in Fig.~\ref{fig:augmented_plant} are arbitrary Gaussian processes that have flat unitary amplitude spectral density.
$\tilde{N}_1(s)$ and $\tilde{N}_2(s)$ are transfer functions used to model the sensor noises $N_1(s)$ and $N_2(s)$ respectively.
The amplitude spectral densities of $N_1(s)$ and $N_2(s)$ are represented by the magnitude of these transfer functions, i.e.
\begin{eqnarray}
	N^\mathrm{ASD}_1(\omega) = \left| \tilde{N}_1(j\omega) \right|\,, \\
	N^\mathrm{ASD}_2(\omega) = \left| \tilde{N}_2(j\omega)  \right|\,,
	\label{eq:asd_model}
\end{eqnarray}
where $N^\mathrm{ASD}_1(\omega)$ and $N^\mathrm{ASD}_2(\omega)$ are the amplitude spectral densities (ASDs) of the sensor noises $N_1(s)$ and $N_2(s)$.
$W_1(s)$ and $W_2(s)$ are pre-compensators, which can be shaped to specify the frequency-dependent specifications of the filtered sensor noises.
Here, the regulator of the plant is $H_1(s)$, which is the complementary filter for filtering the sensor noise $N_1(s)$.
The open-loop transfer matrix of the augmented plant is
\begin{equation}
	\mathbf{P}(s)=
	\left[
	\begin{array}{ccc}
				0		&	\tilde{N}_2(s)W_2(s)		&	1	\\
				\tilde{N}_1(s)W_1(s)	&	-\tilde{N}_2(s)W_2(s)	&	0
	\end{array}
	\right]\,.
\label{eqn:open_loop_transfer_matrix}
\end{equation}
And, the closed-loop transfer function matrix is
\begin{equation}
	\mathbf{G}(s)=
	\left[
	\begin{array}{cc}
		H_1(s)\tilde{N}_1(s)W_1(s) & \left(1-H_1(s)\right)\tilde{N}_2(s)W_2(s)
	\end{array}
	\right]\,.
	\label{eqn:close_loop_transfer_function_matrix}
\end{equation}

Consider the simplest case with $W_1(s)=W_2(s)=1$, and noting that $H_2(s) = 1-H_1(s)$, the output of the augmented plant in Fig.~\ref{fig:augmented_plant} has an amplitude spectral density equal to that of the super sensor noise in Eqn.~\eref{eqn:supersensornoise}.
Using the plant with $W_1(s)=W_2(s)=1$ for $\mathcal{H}_2$ synthesis will give complementary filters that minimize the $\mathcal{H}_2$ norm, which is equivalent to the expected root mean square (RMS) value of the super sensor noise.
This application may not be particularly useful as the RMS of the super sensor noise is usually dominated by the low-frequency intrinsic sensor noise of the inertial sensor.
As a consequence, the $\mathcal{H}_2$ super sensor may not benefit from low noise level of the inertial sensor at high frequency.
While this configuration can be useful for some applications, we seek for optimal complementary filters that can reduce both sensor noises at all frequencies, where the level of sensor noises span a few orders of magnitude.
$\mathcal{H}_\infty$ optimization can be a solution to this problem.
This is because the weights $W_1(s)$ and $W_2(s)$ can be specified as the reciprocal of the frequency-dependent upper bounds of the filtered sensor noises $H_1(s)\tilde{N}_1(s)$ and $H_2(s)\tilde{N}_2(s)$, respectively.
Choosing the weights this way is similar to a standard mixed-sensitivity $\mathcal{H}_\infty$ control problem \cite{skogestad} where the weights can be used to specify the upper bounds of the closed-loop sensitivity functions of a feedback system.
In contrast, the weights do not have special meanings in the context of an $\mathcal{H}_2$ problem.

To see how the weights can be chosen such that the super sensor noise is close to the lower bound at all frequencies, one cost function that would be interesting in particular is
\begin{equation}
	J(H_1(j\omega)) = \max_{\omega}\left(\log{N^\mathrm{ASD}_\mathrm{super}\mleft(\omega;H_1(j\omega)\mright)} - \log N^\mathrm{ASD}_\mathrm{min}(\omega)\right)\,,
	\label{eqn:cost_function}
\end{equation}
where $N^\mathrm{ASD}_\mathrm{super}(\omega;H_1(s))$ is the ASD of the super sensor noise and $N^\mathrm{ASD}_\mathrm{min}(\omega)$ is the ASD of the lower bound of the super sensor noise, defined as
\begin{equation}
	N^\mathrm{ASD}_\mathrm{min}(\omega) \equiv \min\left(N^\mathrm{ASD}_1(\omega),N^\mathrm{ASD}_2(\omega)\right)\,.
\end{equation}
Minimization of the cost function Eqn.~\eref{eqn:cost_function} would give optimal complementary filters $H_1(s)$ and $H_2(s)\equiv 1-H_1(s)$ that best suppress the super sensor noise equally close to the lower bound at all frequencies in logarithmic scale.

To convert this cost function to the objective of the $\mathcal{H}_\infty$ problem, consider the frequency region where $N^\mathrm{ASD}_1(\omega) \gg N^\mathrm{ASD}_2(\omega)$.
In this case, the super sensor noise in Eqn.~\eref{eqn:cost_function} could be approximated by
\begin{equation}
	N^\mathrm{ASD}_\mathrm{super}(\omega)\approx \left| H_1(j\omega)\right| \left|\tilde{N}_1(j\omega)\right|\,,
	\label{eqn:super_sensor_noise_approximate}
\end{equation} 
and the lower bound of the sensor noise is
\begin{equation}
	N^\mathrm{ASD}_\mathrm{min}(\omega) = \left| \tilde{N}_2(j\omega) \right|\,.
	\label{eqn:lower_bound_sensor_noise_approximate}
\end{equation}
Substituting Eqn.~\eref{eqn:super_sensor_noise_approximate} and Eqn.~\eref{eqn:lower_bound_sensor_noise_approximate} into Eqn.~\eref{eqn:cost_function}, we get
\begin{equation}
	J(H_1(s)) \approx \max_\omega \left(\log \frac{\left| H_1(j\omega) \right| \left|\tilde{N}_1(j\omega)\right|}{\left|\tilde{N}_2(j\omega)\right|} \right)\,.
	\label{eqn:cost_function_approximate}
\end{equation}
Minimizing this is equivalent to minimizing a similar cost function without the logarithm, i.e.
\begin{equation}
	J'(H_1(j\omega)) = \max_\omega \left(\frac{\left| H_1(j\omega)\right| \left|\tilde{N}_1(j\omega)\right|}{\left|\tilde{N}_2(j\omega)\right|} \right)\,.
	\label{eqn:cost_function_equivalent}
\end{equation}

Now, let us find pre-compensators $W_1(s)$ and $W_2(s)$ such that the $\mathcal{H}_{\infty}$ synthesis of the generalized plant in Fig.~\ref{fig:augmented_plant} is equivalent as to minimizing the cost function~\eref{eqn:cost_function_equivalent}.
Assuming that $W_1(s)$ and $W_2(s)$ are set such that the magnitude of the first term in Eqn.~\eref{eqn:close_loop_transfer_function_matrix} is much greater than the second term, the $\mathcal{H}_\infty$ norm of the closed-loop transfer function $\mathbf{G}(s)$ could be approximated as
\begin{equation}
	\left\| \mathbf{G}(s) \right\|_\infty \approx \max_\omega \left(\left| H_1(j\omega)\right| \left|\tilde{N}_1(j\omega)\right| \left| W_1(j\omega) \right|\right)\,.
	\label{eqn:h_infinity_norm_approximate}
\end{equation}
Comparing Eqn.~\eref{eqn:cost_function_equivalent} and Eqn.~\eref{eqn:h_infinity_norm_approximate}, it is obvious that if $W_1(s)=1/{\tilde{N}_2(s)}$, the two cost functions become equivalent to each other.
A similar argument can be made for the case $N^\mathrm{ASD}_2(\omega) \gg N^\mathrm{ASD}_1(\omega)$, which gives $W_2(s)=1/{\tilde{N}_1(s)}$.
Therefore, by setting the weighting functions $W_1(s)=1/{\tilde{N}_2(s)}$ and $W_2(s)=1/{\tilde{N}_1(s)}$, the $\mathcal{H}_\infty$ norm of the plant in Fig.~\ref{fig:augmented_plant} is approximately equal to maximum difference between the ASD of the super sensor noise and its lower bound, as described by Eqn.~\eref{eqn:cost_function}.
It follows that $\mathcal{H}_\infty$ synthesis will give an optimal filter $H_1(s)$ while its complementary filter can be obtained from the complementary condition, i.e. $H_2(s) \equiv 1-H_1(s)$

There is another simpler explanation behind these weighting functions $W_1(s)=1/{\tilde{N}_2(s)}$ and $W_2(s)=1/{\tilde{N}_1(s)}$.
Again, the weighting functions $W_1(s)$ and $W_2(s)$ can be thought as inverse frequency-dependent specifications of the sensor noises $N_1(s)$ and $N_2(s)$, respectively~\cite{dehaeze19_compl_filter_shapin_using_synth}.
When $| N_1(j\omega) | > | N_2(j\omega) |$, the target specification for $N_1(s)$ is $N_2(s)$ but not lower.
Any specification lower than that would be over-compensating as the super sensor noise will be dominated by the higher one.
Again, the same argument can be made for $N_2(s)$.
And, it follows that the weighting functions should be set as $W_1(s)=1/{\tilde{N}_2(s)}$ and $W_2(s)=1/{\tilde{N}_1(s)}$, if there are no specific requirements for the sensor noises.

\subsection{Sensor noise modeling\label{sec:sensor_noise_modeling}}
To fully specify the generalized plant for a complementary filter sensor fusion configuration shown in Fig.~\ref{fig:augmented_plant}, $\tilde{N}_1(s)$ and $\tilde{N}_2(s)$ must be specified.
They are the transfer function models whose norms are shaped to fit the amplitude spectral densities of the sensor noises $N^\mathrm{ASD}_1(\omega)$ and $N^\mathrm{ASD}_2(\omega)$, as in Eqn.~\eref{eq:asd_model}.
This problem can be as simple as a curve fitting problem but with one caveat.
That is, the inverse of the transfer function models must be stable, minimum-phase, and proper as their inverse will be invoked.
This restricts the transfer function models to acquire the same number of zeros and poles, all with negative real parts, meaning that they will have flat responses at very low and very high frequencies.
This should not be a problem so long as the features of the sensor noises are modeled within a frequency of interest.

There are a few ways to model noise spectral densities but this is not the main purpose of this paper.
For the completeness of the complementary filter method, a simple but effective method is provided to model the sensor noises.
But, readers are free to model the sensor noises using their own method as long as the method results in transfer function models with norms that fit well with the amplitude spectral densities of the sensor noises in question.
Also note that, as is the case for any $\mathcal{H}_\infty$ problems, the solution to the $\mathcal{H}_\infty$ synthesis problem is only optimal relative to the cost function specified.
This means that the complementary filters synthesized this way will be optimal relative to the modeled sensor noises, but not necessarily the real sensor noises.

We begin with a generic polynomial transfer function model
\begin{equation}
	F(s;a_i,b_i) = \frac{\sum_{i=0}^{n} b_i s^i}{\sum_{i=0}^{n} a_i s^i}\,,
	\label{eqn:transfer_function_model}
\end{equation}
where $a_i$ and $b_i$ are the coefficients of the polynomial and $n$ is the order of the transfer function.
The goal is to find optimal parameters that minimizes a cost function
\begin{equation}
	J_\mathrm{noise}(a_i, b_i) = \sum_{m=1}^M\left(\log \left| F(j\omega_m;a_i, b_i) \right| - \log N(\omega_m)\right)^2\,,
	\label{eqn:cost_function_noise}
\end{equation}
where $m=1,2,3,\ldots,M$, $M$ is the number of data points, $\omega_m$ are the frequency values of the sensor noise data, and $N(\omega_m)$ is the ASD of the sensor noise.
As is mentioned, the transfer function models need to have flat responses at very low and very high frequencies.
It is useful to pad the data with flat lines below and above the measurement frequencies.

To minimize the cost function in Eqn.~\eref{eqn:cost_function_noise}, local minimization methods are recommended because the parameters $a_i$ and $b_i$ are usually not well bounded.
For the same reason, it is recommended to replace $a_i$ and $b_i$ with $10^{\log{a_i}}$ and $10^{\log{b_i}}$ in Eqn.~\eref{eqn:transfer_function_model} and optimize $\log{a_i}$ and $\log{b_i}$ instead, as the parameters could vary with large orders of magnitude.
Local minimization methods, such as the Nelder-Mead method~\cite{nelder_mead} and Powell's method~\cite{powell}, require initial specifications of the parameters, which can be hard to obtain.

As an intermediate step, consider a zero-pole-gain (ZPK) model
\begin{equation}
	F_\mathrm{ZPK}(s;z_i, p_i,k) = k\frac{\prod_{i=1}^{n}s-z_i}{\prod_{i=1}^{n}s-p_i}\,,
\end{equation}
where $z_i$ and $p_i$ are negative real-valued zeros and poles of the transfer function, and $k$ is the gain of the transfer function.
These zeros and poles are corner-frequencies where the amplitude frequency response changes slope by 20 decibels per decade.
They can be easily added and tuned manually to lay out the general shape of the model that fits the sensor noise data.
Alternatively, it is possible to replace $F(s;a_i,b_i)$ in Eqn.~\eref{eqn:cost_function_noise} with the ZPK model $F_\mathrm{ZPK}(s;z_i,p_i,k)$ and use global optimization methods, such as differential evolution~\cite{differential_evolution}, to find the zeros and poles.
This is possible since the zeros and poles are expected to be bounded within frequency space of the measured sensor noise data.
Again, it is recommended to fit $\log{z_i}$ and $\log{p_i}$ instead due to their large dynamic range.
After obtaining a ZPK model, it can be expanded to the polynomial form to obtain the initial coefficients for the transfer function model Eqn.~\eref{eqn:transfer_function_model} used for the local minimization of Eqn.~\eref{eqn:cost_function_noise}.
At last, all non-negative real parts of the zeros and poles must be negated to obtain the final stable, minimum-phase, proper transfer function that fits the ASD of the sensor noise.

The choice of the transfer function order $n$ during the ZPK fitting depends on the frequency dependency of the noise profile.
In general, this order needs to be higher than the most significant frequency dependency of the noise.
For example, if the ASD of the sensor noise has a $1/f^{-3.5}$ dependency, a choice of $n=4$ would be a reasonable (and often sufficient) choice  for initialization.
Increasing $n$ would necessarily lead to a better fit of the noise spectrum.
However, in practice, it was shown a choice of an excessive $n$ would lead to pole-zero cancellation at irrelevant frequencies during the transfer function fit, which are useless features for the model.
This is typically a good termination point when trying with different $n$.

	\section{Results\label{sec:result}}
In this section, the proposed $\mathcal{H}_\infty$ method is used to synthesize three different pairs of sensors.
The sensors to be considered are linear variable differential transformers (LVDTs) and geophones, which are commonly used in active isolation systems in current gravitational-wave detectors.
The three configurations are
\begin{enumerate}
	\item LVDT and geophone with sensor noises estimated from Ref.~\cite{Sekiguchi:2016cdw}.
	\item LVDT with seismic noise coupling and geophone.
	\item Hypothetical LVDT-like and geophone-like sensors.
\end{enumerate}
Again, the $\mathcal{H}_\infty$ complementary filters are obtained by $\mathcal{H}_\infty$ synthesis, which seek an optimal filter $H_1(s)$ that optimized the $\mathcal{H}_\infty$ norm, Eqn.~\eref{eqn:h_infinity_norm}, of the closed loop transfer matrix defined by Eqn.~\eref{eqn:close_loop_transfer_function_matrix}.

\subsection{Sensor fusion of a relative displacement sensor and an inertial Sensor \label{sec:lvdt_geophone}}
In this section, the proposed method will be demonstrated by synthesizing complementary filters for blending an LVDT and a geophone, referred to as sensor 1 and sensor 2, respectively.
The sensor noises are taken from Figure 5.8 in Ref.~\cite{Sekiguchi:2016cdw}, and can be well described by Eqn.~\eref{eqn:sensor_noise_quadsum}.
\begin{equation}
	N^\mathrm{ASD}(f; n_a, n_b, a, b) = \left[\left(\frac{n_a}{f^a}\right)^2 + \left(\frac{n_b}{f^b}\right)^2 \right]^{\frac{1}{2}}\,\frac{\mu \mathrm{m}}{\sqrt{\mathrm{Hz}}}\,,
	\label{eqn:sensor_noise_quadsum}
\end{equation}
where $f$ is frequency in Hz, $n_a$, $n_b$, $a$, and $b$ are some parameters of the model.
For demonstration, the parameters are chosen by a graphical estimation from a figure shown in Ref.~\cite{Sekiguchi:2016cdw}.
The parameters for the two sensors are summarized in table.~\ref{table:sensor_noises}.

\begin{table}[!h]
	\caption{\label{table:sensor_noises}LVDT and geophone intrinsic noise parameters.}
	\begin{indented}
		\item[]\begin{tabular}{@{}lllll}
			\br
			Sensor & $n_a$ & $n_b$ & $a$ & $b$ \\
			\mr
			LVDT & $10^{-2.07}$ & $10^{-2.3}$ & 0.5 & 0\\
			Geophone & $10^{-5.46}$  & $10^{-5.23}$ & 3.5 & 1\\
			\br
		\end{tabular} 
	\end{indented}
\end{table}

\begin{figure}[!h]
	\centering
	\includegraphics[width=0.7\linewidth]{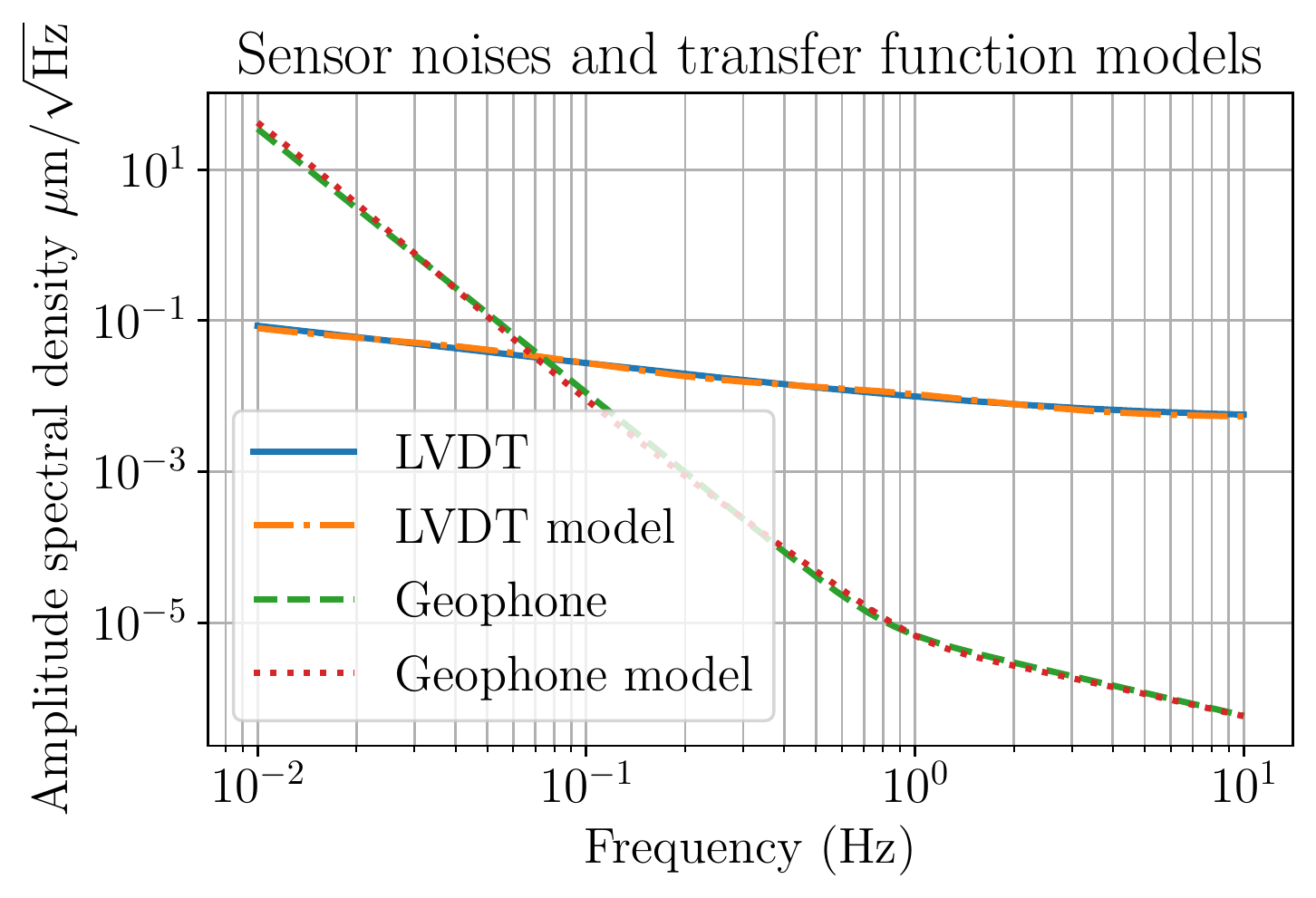}
	\caption{Amplitude spectral densities of the sensor noises and the transfer function models. Blue solid:  LVDT intrinsic noise. Orange dash-dot: transfer function model of the LVDT noise. Green dashed: geophone intrinsic noise. Red dotted: transfer function model of geophone noise.}
	\label{fig:sensornoiselvdtgeophone}
\end{figure}

The amplitude spectral densities of the sensor noises and their transfer function models are shown in Fig.~\ref{fig:sensornoiselvdtgeophone}.
In this case, sensor noise 1 is fitted with a $3^\mathrm{rd}$-order transfer function while sensor noise 2 is fitted with a $4^\mathrm{th}$-order one. 
The selection of the two transfer function orders is based of the exponents $a$ and $b$ of each sensor noise profile as shown in table.~\ref{table:sensor_noises}.
The order are chosen to be the smallest integer that is larger than both $a$ and $b$.

As shown in Fig.~\ref{fig:sensornoiselvdtgeophone}, the amplitude responses of the transfer functions fit well to the ASD of the sensor noises.

\begin{figure}[!h]
	\centering
	\includegraphics[width=0.7\linewidth]{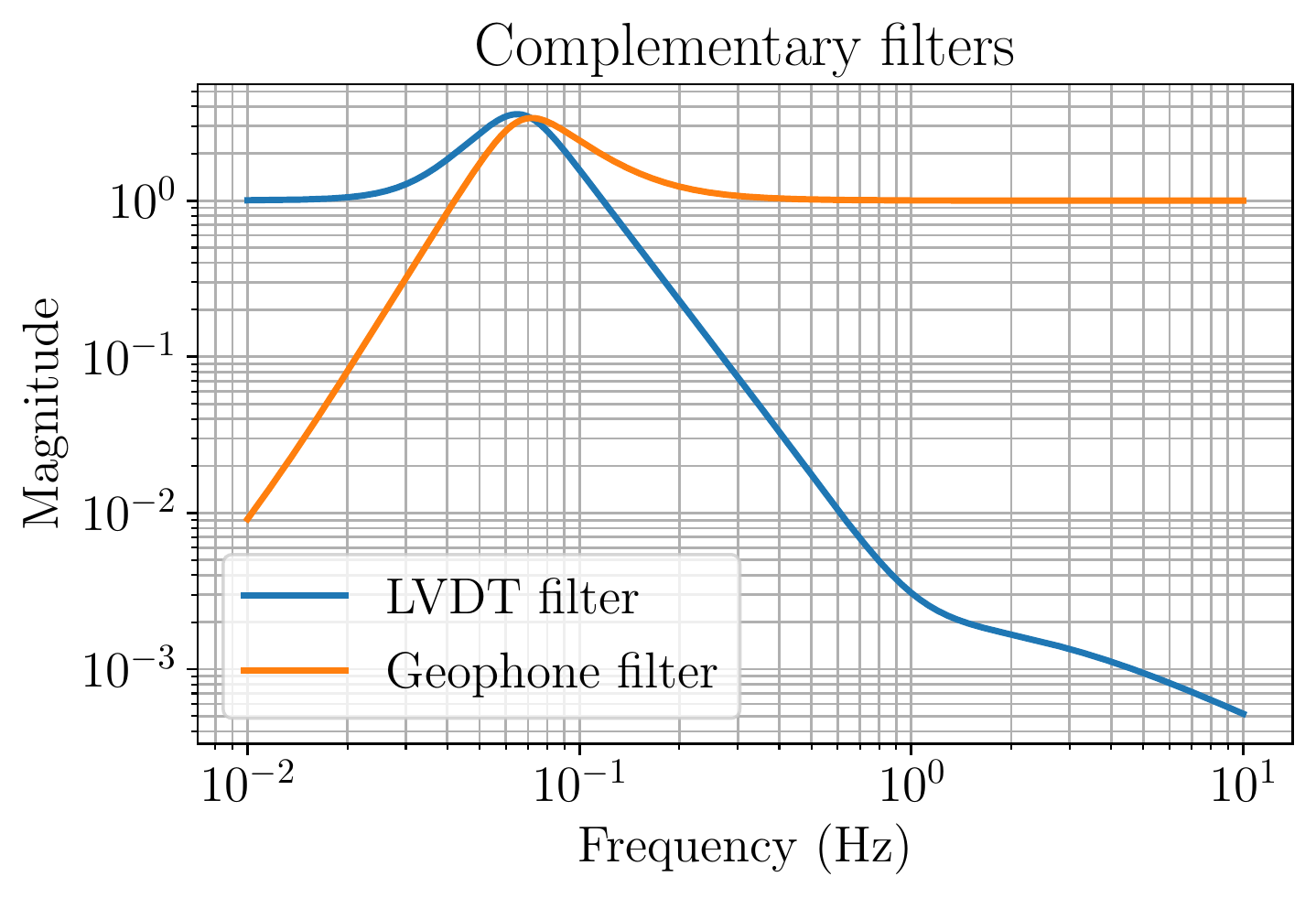}
	\caption{Complementary filters. Blue: low-pass filter for the LVDT. Orange: high-pass filter for the geophone.}
	\label{fig:hinfcomplementaryfilterlvdtgeophone}
\end{figure}

The transfer functions are used to synthesize complementary filters according to the proposed method.
The complementary filters synthesized using $\mathcal{H}_\infty$ method are shown in Fig.~\ref{fig:hinfcomplementaryfilterlvdtgeophone}.

\begin{figure}[!h]
	\centering
	\includegraphics[width=0.7\linewidth]{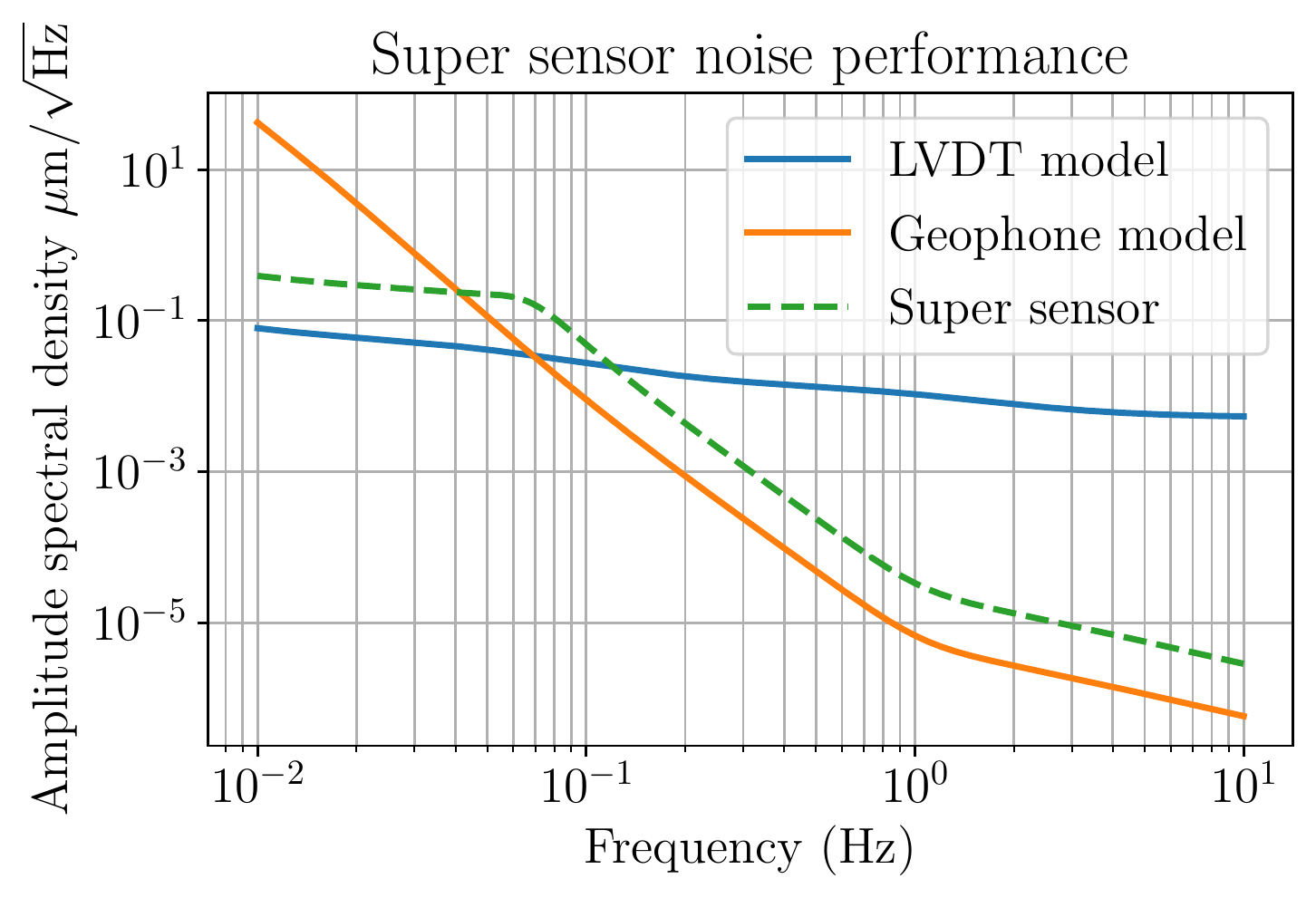}
	\caption{Amplitude spectral densities of the sensor noises. Blue solid: LVDT noise model. Orange: geophone noise model. Green dashed: predicted super sensor noise.}
	\label{fig:supersensornoiselvdtgeophone}
\end{figure}

Using the synthesized filters $H_1(s)$ and $H_2(s)\equiv 1-H_1(s)$, the amplitude spectral density of super sensor noise is predicted by
\begin{equation}
	N^\mathrm{ASD}_\mathrm{super}(\omega) = \left[\left| H_1(j\omega) N^\mathrm{ASD}_1(\omega)\right|^2 + \left| H_2(j\omega) N^\mathrm{ASD}_2(\omega)\right|^2 \right]^\frac{1}{2}\,.
\end{equation}
The predicted ASD of the super sensor noise is shown in Fig.~\ref{fig:supersensornoiselvdtgeophone}.
As can be seen, the super sensor noise is equally close to the lower bound in logarithmic sense at all frequencies, as expected.

\subsection{Sensor fusion for a seismic-noise-coupled relative displacement sensor and an inertial sensor}
In this section, the sensor fusion of a seismic-noise-coupled displacement sensor (LVDT) and an inertial sensor (geophone) is demonstrated and compared.

LVDTs are relative displacement sensors.
When they are used on the first stage of an active isolation platform, they read relative displacements between the suspended platform and the ground.
The ground motion in the LVDT readout is an unwanted signal for active isolation.
Therefore, the seismic noise is often considered a part of the LVDT noise.
The seismic noise features a peak around 0.1-0.5 Hz, which correspond to the secondary microseisms.
If this is not filtered, the microseismic disturbance cannot be actively attenuated.
Or, in the worst case, the seismic noise will be injected to the isolation platform.
Pre-designed complementary filters lack a quality that effectively suppresses the microseism while the $\mathcal{H}_\infty$ method can take the microseismic peak into account and optimize filters that can better attenuate the seismic noise.
To take seismic noise into account, the LVDT noise model is reused from Sec.~\ref{sec:lvdt_geophone} but is multiplied by a transfer function:
\begin{equation}
	N_\mathrm{seis}(s) = \prod_{i=1}^4\frac{ \frac{1}{\omega_i^2}s^2 + \frac{1}{\omega_i}s + 1}{ \frac{1}{\omega_i^2}s^2 + \frac{1}{\omega_iq_i}s + 1} \,,
\end{equation}
where $\omega_i =2\pi\times\{0.15, 0.2, 0.25, 0.3\}\, \rm{rad}/s$, and $q_i=3$.
This will simulate a microseismic peak around 0.2 Hz.

\begin{figure}[!h]
	\centering
	\includegraphics[width=0.7\linewidth]{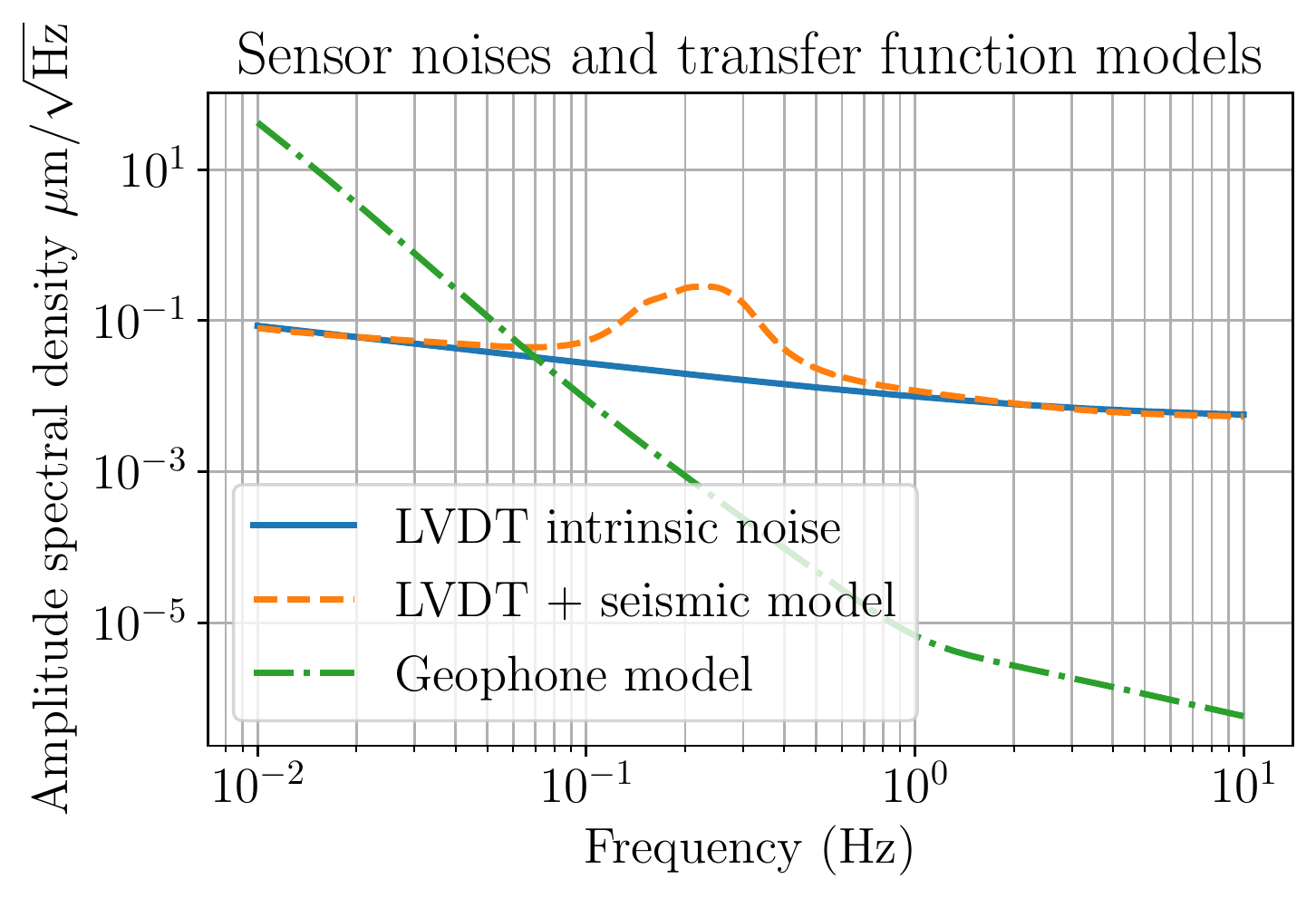}
	\caption{Amplitude spectral densities of the sensor noises and the transfer function models. Blue solid: Original intrinsic LVDT noise. Orange dashed: transfer function model of the seismic-noise coupled LVDT sensor noise. Green dash-dot: transfer function model of the geophone noise.}
	\label{fig:sensornoiselvdtmicroseismgeophone}
\end{figure}

\begin{figure}[!h]
	\centering
	\includegraphics[width=0.7\linewidth]{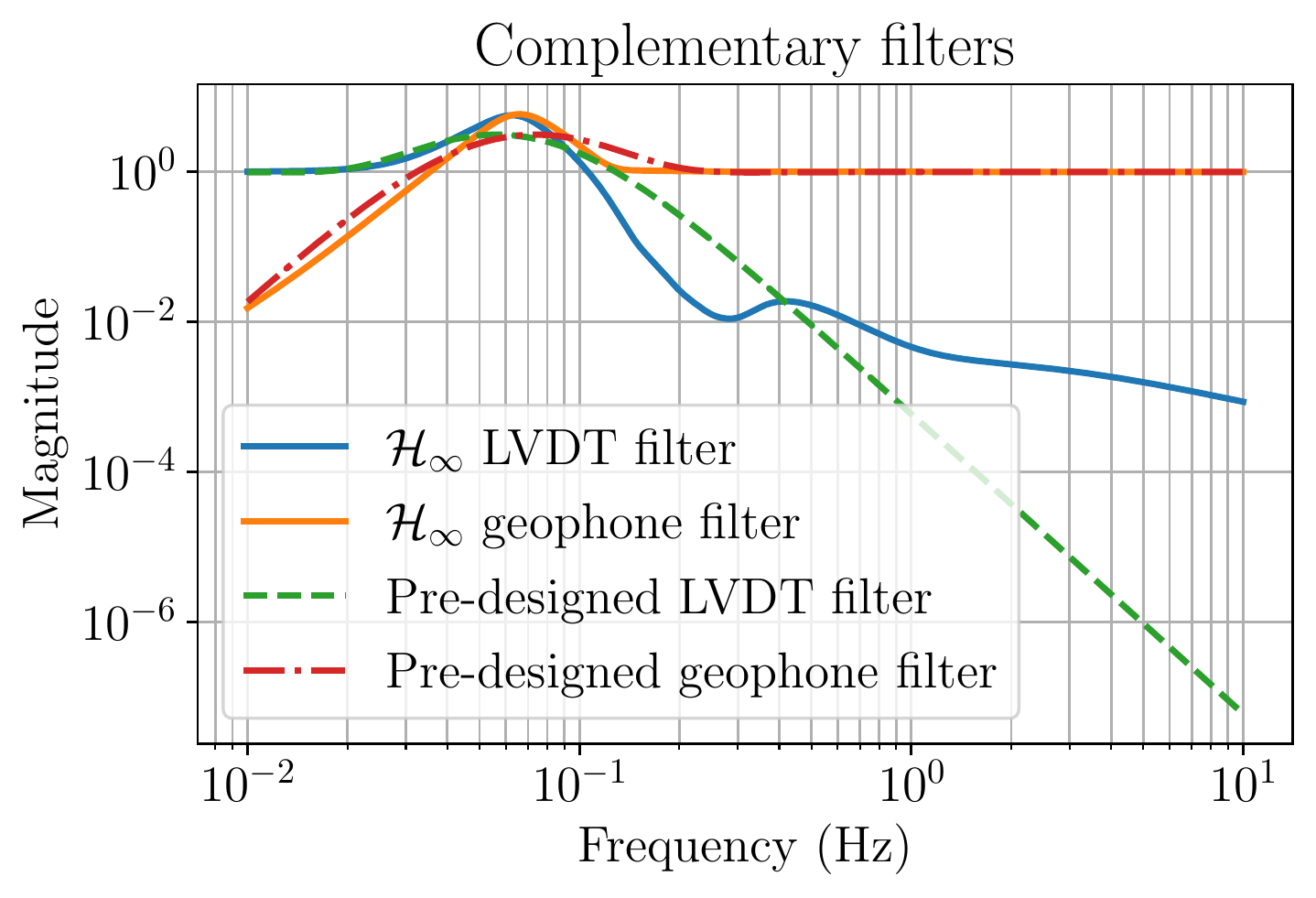}
	\caption{Complementary filters. Blue solid: low-pass filter for seismic noise-coupled LVDT, synthesized using $\mathcal{H}_\infty$ method.  Orange solid: high-pass filter for geophone, synthesized using $\mathcal{H}_\infty$ method. Green dashed: pre-designed low-pass filter for the seismic noise-coupled LVDT. Red dashed: pre-designed high-pass filter for the geophone.}
	\label{fig:hinfcomplementaryfilterlvdtmicroseismgeophone}
\end{figure}

The other sensor to be blended with this LVDT sensor is the geophone used in Sec.~\ref{sec:lvdt_geophone}.
The sensor noise models are shown in Fig.~\ref{fig:sensornoiselvdtmicroseismgeophone}.
The transfer function models of the seismic-noise-coupled LVDT noise and the geophone noise are used to synthesize optimal complementary filters shown in Fig.~\ref{fig:hinfcomplementaryfilterlvdtmicroseismgeophone} (Blue and orange solid lines).

The pre-designed complementary filters to be compared with are $7^\mathrm{th}$-order filters with $4^\mathrm{th}$-order roll-off.
These filters were pre-designed specifically to use with this sensor fusion configuration in Ref.~\cite{Sekiguchi:2016cdw}.
The filters were chosen to suppress the inertial sensing noise, which has a frequency dependency of $f^{-3.5}$ at low frequency.
There is only one design parameter for the pre-designed filter, that is, the blending frequency.
The blending frequency in this case is chosen to be at the crossover frequency between the LVDT and geophone noise, as advised in Ref.~\cite{Sekiguchi:2016cdw}.
In this case, the blending frequency of the pre-designed filters is  64.47 mHz.
The pre-designed complementary filters are also shown in Fig.~\ref{fig:hinfcomplementaryfilterlvdtmicroseismgeophone} (greed dashed and red dash-dot lines).

As can be seen, the low-pass filter (blue solid line in Fig.~\ref{fig:hinfcomplementaryfilterlvdtmicroseismgeophone})) generated using the proposed $\mathcal{H}_\infty$ method has a notch feature around 0.1-0.3 Hz.
Compared to the pre-designed filter, this provides significantly higher seismic noise attenuation around the microseismic frequency.
Conventionally, the additional notch features in the low-pass filter were artificially added.
Example filters can be found in Refs.~\cite{Matichard:2015eva,lucia_thesis}.
In contrast, the notch feature is a natural result of the $\mathcal{H}_\infty$ optimization.

\begin{figure}[!h]
	\centering
	\includegraphics[width=0.7\linewidth]{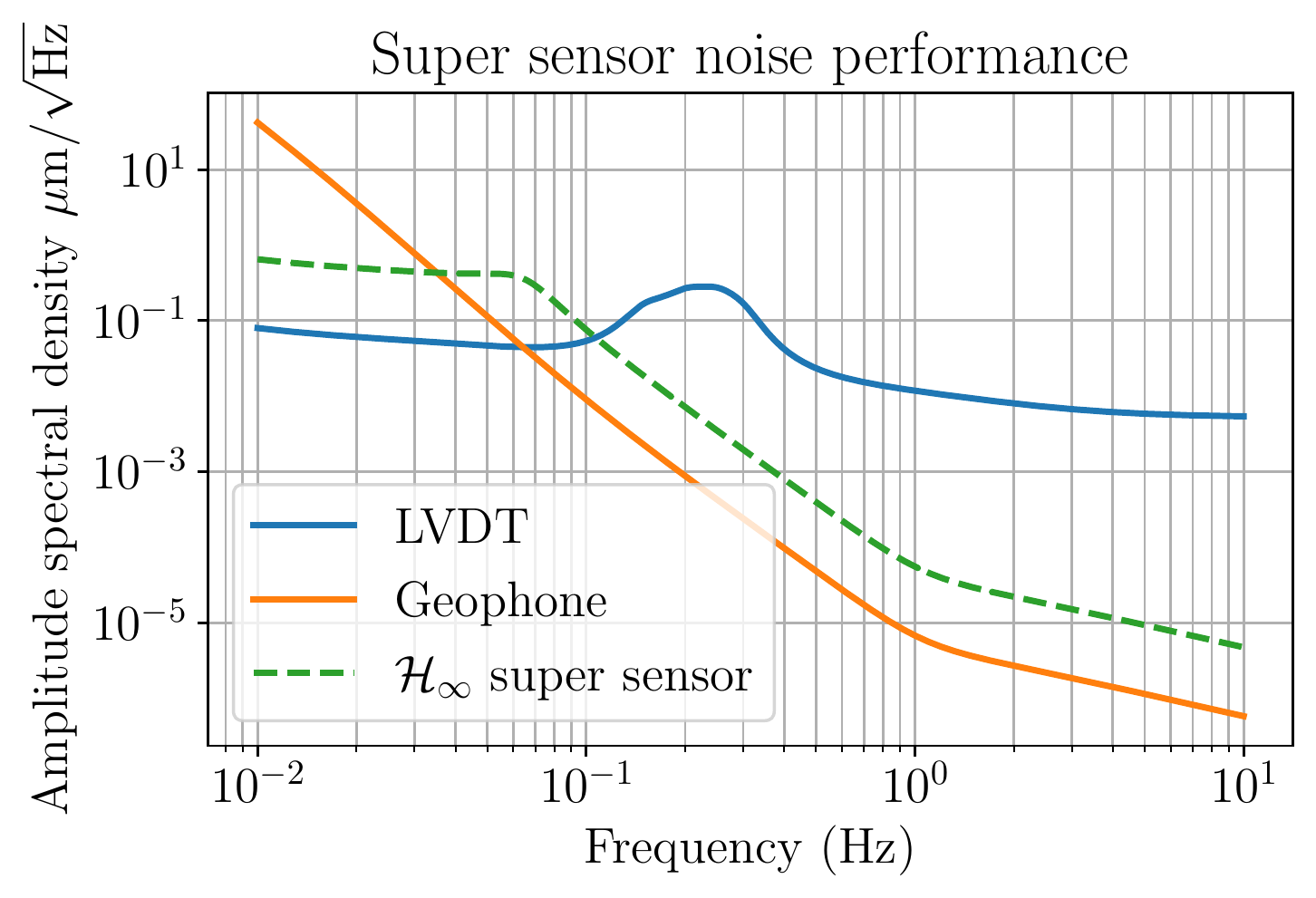}
	\caption{Sensor sensor noise. Blue solid: seismic-noise-coupled LVDT noise model. Orange solid: geophone noise model. Green dashed: super sensor with $\mathcal{H}_\infty$-optimal complementary filters.}
	\label{fig:supersensornoiselvdtmicroseismgeophone}
\end{figure}

Fig.~\ref{fig:supersensornoiselvdtmicroseismgeophone} shows the predicted super sensor noise performance using the $\mathcal{H}_\infty$-optimal complementary filters..
As can be seen, the noise of the $\mathcal{H}_\infty$ super sensor kept an amplitude spectral density close to the lower bound at all frequencies, including the frequency range of the seismic noise peak.

\begin{figure}[!h]
	\centering
	\includegraphics[width=0.7\linewidth]{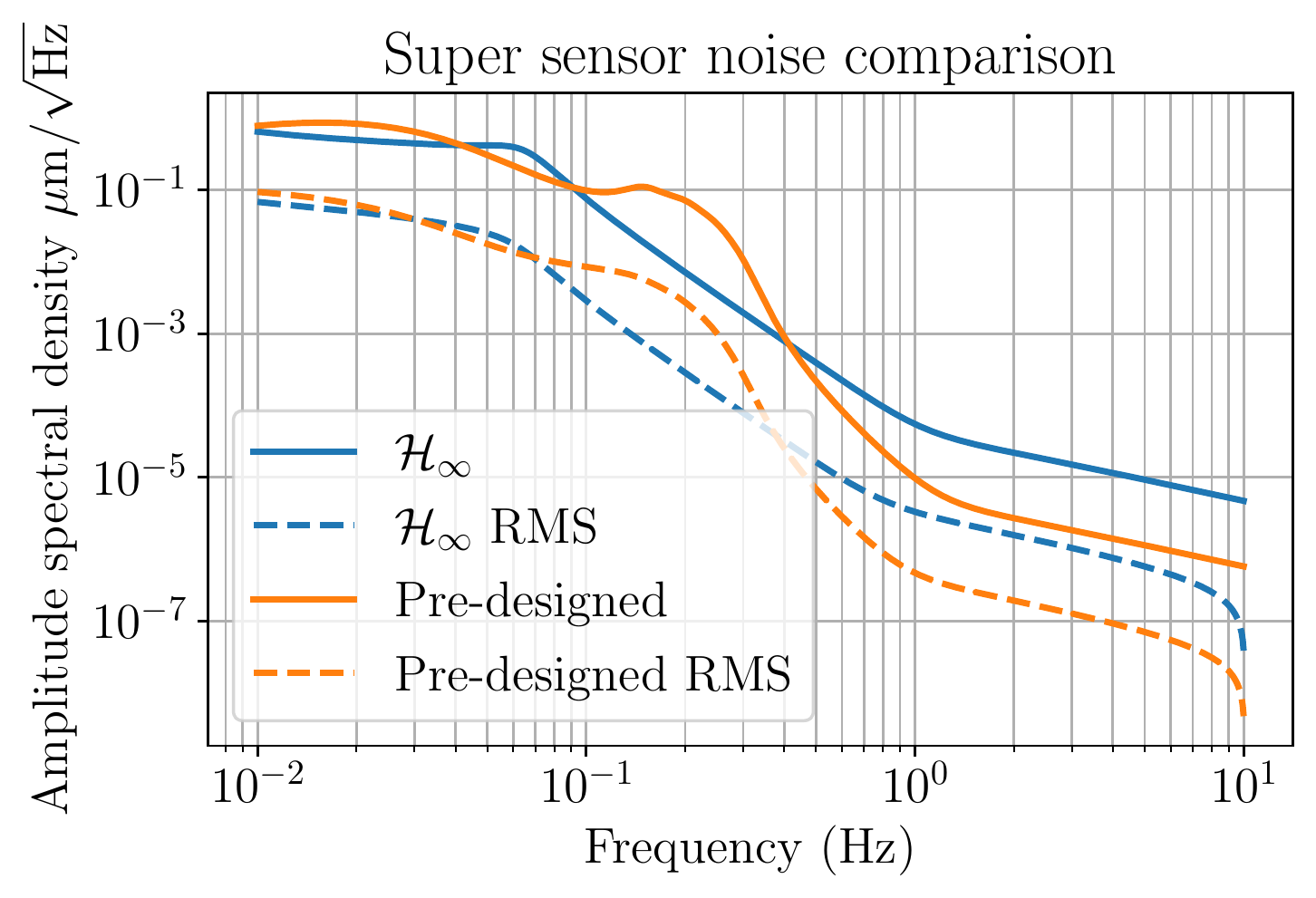}
	\caption{Sensor sensor noise comparison. Blue solid: noise of the super sensor using $\mathcal{H}_\infty$ complementary filters. Blue dashed: expected RMS value of the $\mathcal{H}_\infty$ super sensor noise. Orange solid: noise of the super sensor using pre-designed complementary filters. Orange dashed: expected RMS value of the pre-designed super sensor.}
	\label{fig:supersensornoisecomparisonlvdtmicroseismgeophone}
\end{figure}

In Fig.~\ref{fig:supersensornoisecomparisonlvdtmicroseismgeophone}, the noise performances of the super sensor using $\mathcal{H}_\infty$ filters and the pre-designed filters are compared.
The expected RMS values of the sensor noises are for comparison and it is defined as
\begin{equation}
	N_\mathrm{RMS}(f) = \left[\int^\infty_f\,N^\mathrm{ASD}(f')^2\,df'\right]^\frac{1}{2}\,,
\end{equation}
where $N^\mathrm{ASD}(f)$ is the amplitude spectral density of the sensor noise.
The expected RMS value of each super sensor noise is shown as blue dashed line and orange dashed line in Fig.~\ref{fig:supersensornoisecomparisonlvdtmicroseismgeophone}.
In this case, the expected RMS integrated from 10 Hz to 0.01 Hz.
The expected RMS of the $\mathcal{H}_\infty$ and pre-designed super sensors are 0.06836 $\mu\rm{m}$ and 0.09403 $\mu\rm{m}$, respectively.
Note that these values are one type of performance indexes only and no definitive conclusions should be made as long as they fall into the same order of magnitude.
Moreover, the cost function of the $\mathcal{H}_\infty$ optimization is not necessarily related to the expected RMS.
Therefore, there is no guarantee that the $\mathcal{H}_\infty$ super sensors will always have a lower noise level in terms of the expected RMS.
To minimize the expected RMS, $\mathcal{H}_2$ synthesis could be used instead but this is not the purpose of this paper.

One way to evaluate the seismic attenuation performance of the sensor sensors would be comparing the suppression ratio between the original LVDT noise and the super sensor noises at the microseismic peak.
In this particular example, the peak of the LVDT noise spectrum is located at 0.231 Hz and the noise level is 0.2829 $\mu\rm{m}/\sqrt{\rm{Hz}}$.
At the microseismic peak, the ASDs of the $\mathcal{H}_\infty$ super sensor noise and the pre-designed super sensor noise read 0.004499 $\mu\rm{m}/\sqrt{\rm{Hz}}$ and 0.04614 $\mu\rm{m}/\sqrt{\rm{Hz}}$, respectively.
They offer a suppression ratio of 62.89 and 6.131 for the $\mathcal{H}_\infty$ case and the pre-designed case, respectively. This means the $\mathcal{H}_\infty$ filters provide more than an order of magnitude attenuation of microseismic noise compared to the pre-designed filters.

\FloatBarrier

\subsection{Sensor fusion of hypothetical relative displacement sensor and inertial sensors}
There are many types of sensors that can be used to achieve sensor fusion in active isolation systems.
They all contain different sensor noise profiles and undoubtedly would require different sets of complementary filters if they are used in a sensor fusion configuration.
Generally, reusing complementary filters from another sensor configuration would lead to sub-optimal performance or even lead to unnecessary noise amplification.
Therefore, complementary filters must be redesigned for new sensor configurations and the proposed $\mathcal{H}_\infty$ method provides a convenient way to do so.

To exemplify this, two new sensor noise profiles are considered in this section.
The ASD of the two sensor noise profiles are simply
\begin{equation}
	N_1^\mathrm{ASD}(f) = 1\,\frac{\mu\mathrm{m}}{\sqrt{\mathrm{Hz}}}
\end{equation}
and
\begin{equation}
	N_2^\mathrm{ASD}(f) = \frac{0.1}{f}\,\frac{\mu\mathrm{m}}{\sqrt{\mathrm{Hz}}}\,.
\end{equation}

\begin{figure}[!h]
	\centering
	\includegraphics[width=0.7\linewidth]{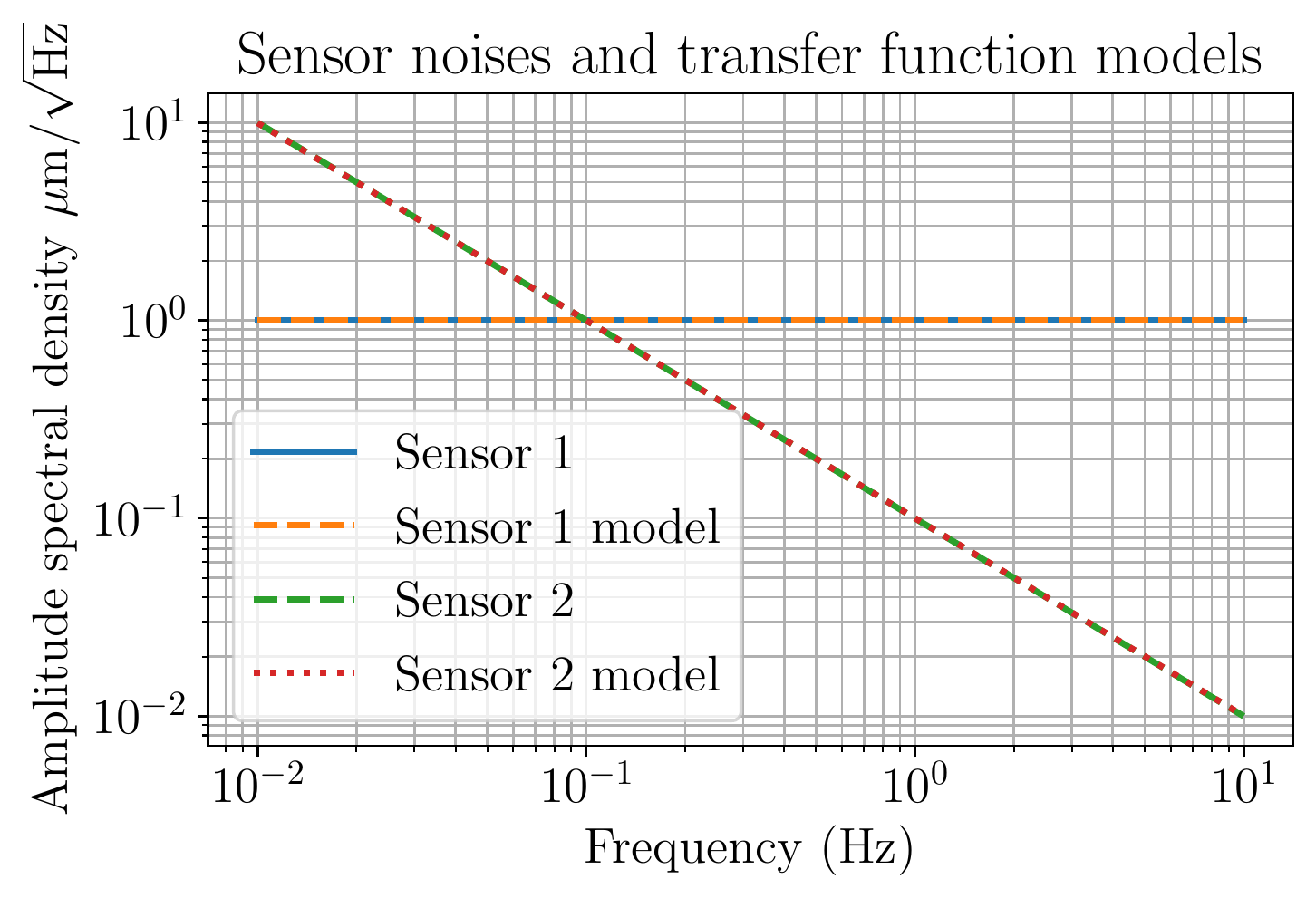}
	\caption{Sensor noise. Blue solid: sensor 1. Orange dash-dot: transfer function model of sensor 1 noise. Green dashed: sensor 2. Red dotted: transfer function model of sensor 2.}
	\label{fig:sensornoisehypothetical}
\end{figure}
Fig.~\ref{fig:sensornoisehypothetical} shows the amplitude spectral densities of the hypothetical sensor noises.

\begin{figure}[!h]
	\centering
	\includegraphics[width=0.7\linewidth]{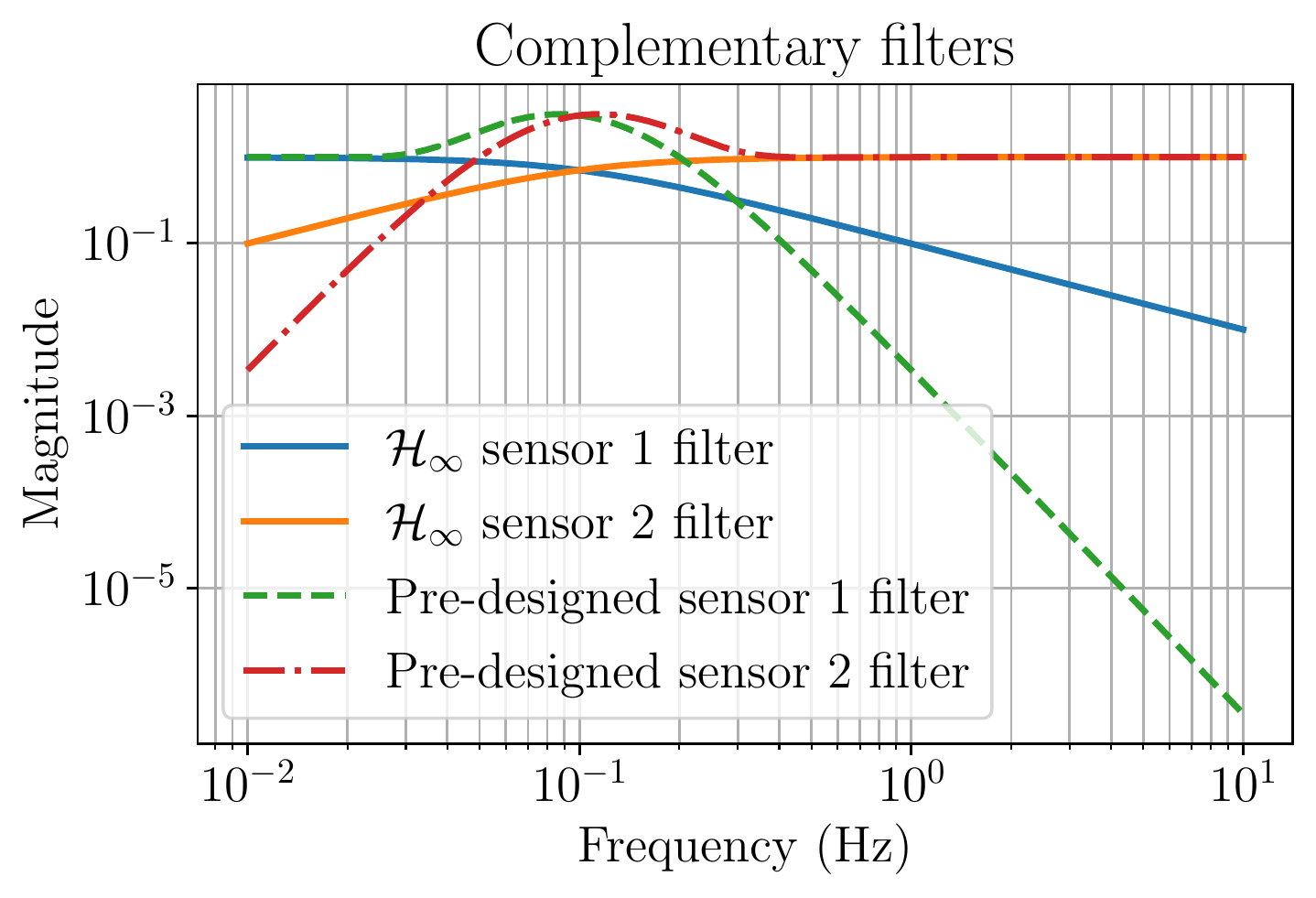}
	\caption{Complementary filters. Blue solid: low-pass filter for sensor 1, synthesized using $\mathcal{H}_\infty$ method.  Orange solid: high-pass filter for sensor 2, synthesized using $\mathcal{H}_\infty$ method. Green dashed: pre-designed low-pass filter. Red dashed: pre-designed high-pass filter.}
	\label{fig:hinfcomplementaryfilterhypothetical}
\end{figure}

Complementary filters were synthesized using the proposed $\mathcal{H}_\infty$ method and they are shown in Fig.~\ref{fig:hinfcomplementaryfilterhypothetical} together with the pre-designed filters.
The pre-designed filters have the same shape as those in Fig.~\ref{fig:hinfcomplementaryfilterlvdtmicroseismgeophone} but with a new blending frequency at 0.1 Hz, i.e. where the two sensor noises meet.
As shown in the figure, the $\mathcal{H}_\infty$-optimal filters have a milder roll-off and with no noise amplification around the blending frequency.

\begin{figure}[!h]
	\centering
	\includegraphics[width=0.7\linewidth]{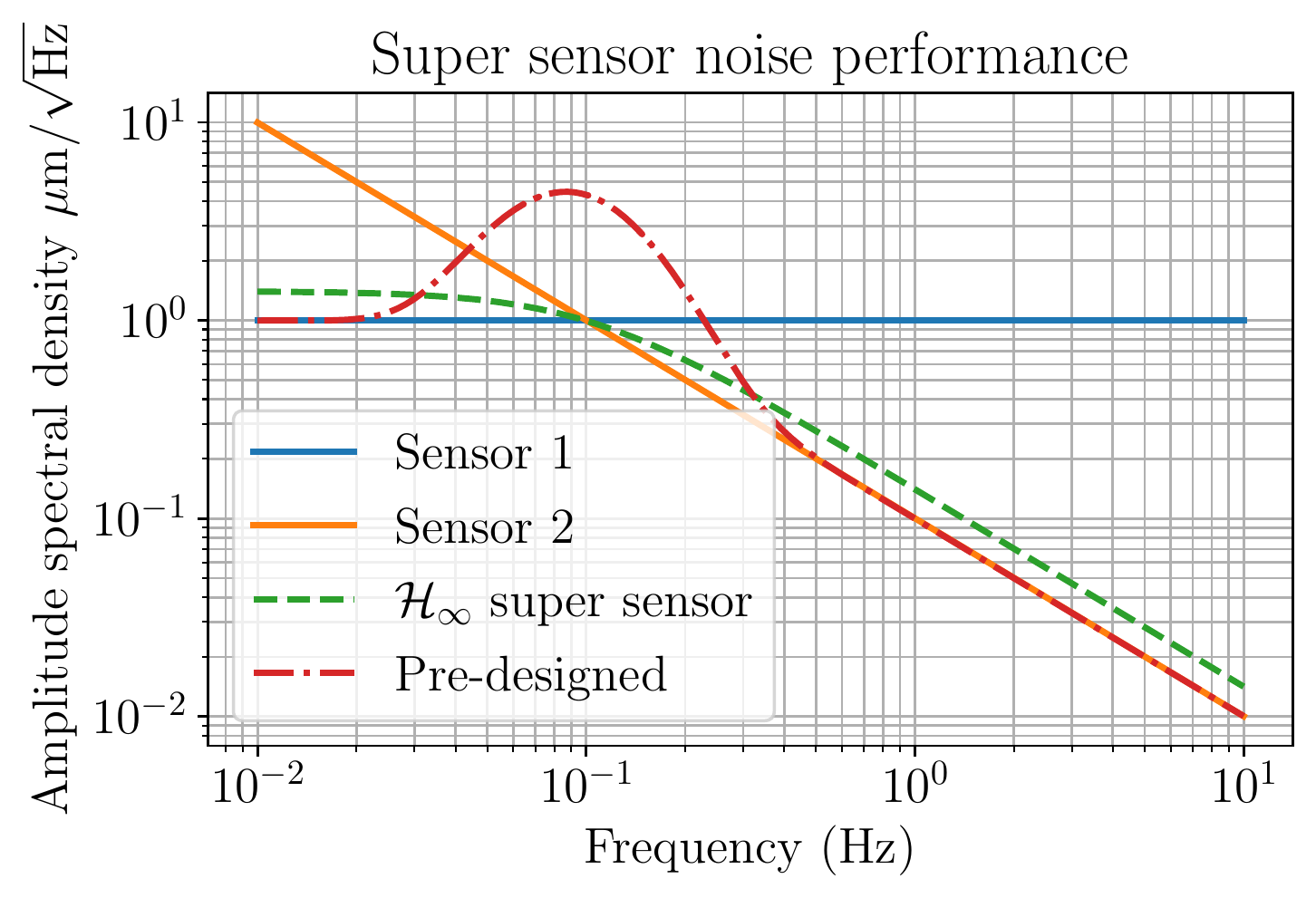}
	\caption{Sensor sensor noise. Green dashed: super sensor with $\mathcal{H}_\infty$-optimal complementary filters. Red dash-dot: super sensor with pre-designed complementary filters.}
	\label{fig:supersensornoisehypothetical}
\end{figure}

The predicted noise performances of the super sensors are shown in Fig.~\ref{fig:supersensornoisehypothetical}.
As can be seen, the super sensor fused with $\mathcal{H}_\infty$-optimal filter has sensor noise almost indistinguishable from the lower bound at all frequencies.
Meanwhile, the super sensor fused with the pre-designed filter has a noise peak of 4.457 $\mu\rm{m}/\sqrt{Hz}$ at around the blending frequency at 0.1 Hz, amplifying the noise.
This is a result of over-suppression at lower and higher frequencies, which is one problem that the $\mathcal{H}_\infty$ method aims to avoid.
This makes the pre-designed filter relatively unsuitable for this particular sensor configuration.
In comparison, the $\mathcal{H}_\infty$ super sensor has a maximum noise amplification of 1.397 times, and the maximum ASD of the noise is 1.397 $\mu \rm{m}/\sqrt{\rm{Hz}}$ at 0.01 Hz.


	\section{Discussions\label{sec:discussion}}
Using the proposed method, it is possible to synthesize optimal complementary filters that best suppress the super sensor noise equally close to the lower bound at all frequencies in logarithmic scale.
Unlike pre-designed filters, the method allows one to make complementary filters that work with any arbitrary sensor, as long as the sensor noise can be modeled.
Optimal complementary filters can be generated from only the sensor noises themselves.
In all the results shown, the complementary filters generated from $\mathcal{H}_\infty$ optimization performed better, compared to the pre-designed ones.
And, the $\mathcal{H}_\infty$ method provides a new way of optimizing complementary filters for virtually any type of sensor fusion configurations. 
This will be extremely useful for future detectors such as the Einstein Telescope.

The method is able to produce complementary filters with special features, such as notches, to cope with the special noise characteristics in sensors.
In a conventional filter shaping process, these special features would need to be added manually by control experts based on practical experience.
This would make the filters arguably sub-optimal, and most importantly, not reproducible.
But, with the proposed method, the features automatically appear in a natural way as a result of optimization.
None or little human intervention is required in the whole synthesis process.
Also because of this, this opens up the possibility of rolling-update of control filters where filters are synthesized automatically in real-time according to changes in the environment, e.g. changes in seismic noise.

To use the $\mathcal{H}_\infty$ method, one requires to model the frequency content of the sensor noises as transfer functions.
For other methods, an empirical model may suffice.
Modeling noise spectrum with transfer functions can be difficult as the amplitude spectrum of sensor noises typically has fractional frequency-dependency, such as $1/f^{3.5}$.
This is not well represented by transfer functions since the transfer functions typically have integer order corresponding to the number of zeros and poles of the underlying system.
One needs to use a higher-order transfer function to represent a sensor noise that has a lower-order fractional frequency-dependency, as is done in Sec.~\ref{sec:lvdt_geophone}.
As a consequence, the noise models fluctuate with a small magnitude around the sensor noises as shown in Fig.~\ref{fig:hinfcomplementaryfilterlvdtgeophone}.
It should be noted that the error in modeling is small compared to the relative difference between the two sensor noises.
This type of error in modeling may not be significant as the important quantity in the cost function is the relative difference between the noises.
In this case, the super sensor would be sub-optimal in the sense that the super sensor noise fluctuates, with a small magnitude, around the truly optimal one.
Another source of modeling error is measurement error, i.e. the model is fitted to a measurement that does not capture the real frequency content of the sensor noise.
This error is common to all methods.
This is not an exclusive problem of the $\mathcal{H}_\infty$ problem and must be solved independently as a modeling problem.
The error in modeling may not have a significant effect on the design of the feedback controller for active isolation since they are separate processes.
The real super sensor noise can always be estimated or measured so the controller designer can design controllers around the real data.

There are a few problems that need to be addressed for practical implementation of complementary filters.
These problems are not necessarily exclusive to the $\mathcal{H}_\infty$ approach, but will cause degradation to the $\mathcal{H}_\infty$ filters, making them less than optimal.
The inter-calibration and alignment between the sensors need to be done well, or else the super sensor response will not be unity.
Moreover, the inertial sensors often require a calibration filter that represents their inverse dynamics.
Mismodeling of the inverse dynamics leads to frequency-dependent calibration mismatch between the inertial sensors and other sensors, which will also cause the super sensor response to be distorted.
The inertial sensor readouts also require substantial prefiltering at an early stage in the control signal path to avoid overflow at low frequency due to integration.
The prefilters add additional attenuation on top of one of the complementary filters, effectively making the complementary filters not complementary.
All of these could cause spurious responses in the control system that could lead to limited control performance or even lead to instability.
A post processing treatment of the complementary filters to account for prefilters is proposed in Ref.~\cite{Heijningen2018TurnUT}.
However, the treatment could ruin the result of the $\mathcal{H}_\infty$ optimization and hence is not preferred.
A future paper will focus on the practical implementation of the $\mathcal{H}_\infty$ complementary filters along with the experimental results.

The proposed $\mathcal{H}_\infty$ method complements some existing control strategies in active isolation systems in current gravitational-wave detectors.
A control strategy was used in LIGO called ``earthquake mode'' where a pre-designed filter is swapped with another one when there is an anticipated earthquake that could cause a lock-loss of the interferometer~\cite{LIGO:2020pzq,ligo_earthquake_mode}.
These filters were made without any information of the upcoming earthquake or seismic noise, which means the filters could again be sub-optimal.
But with the proposed method, the filters can be synthesized in real-time, making the active isolation truly adaptive to the environment and ultimately increasing the duty cycle of the detector.

Active isolation systems in GW detectors use many control filters other than complementary filters that to achieve active alignment control and seismic isolation.
Some examples would be the sensor correction filter, the seismometer feedforward filter, and the feedback controller, which are all used in LIGO~\cite{Matichard:2015eva, pso_sensor_correction}.
Sensor correction filter is a filter used in LIGO that applies on a seismometer, which is used to remove seismic noise coupling from relative displacement sensors, making them available for seismic isolation via feedback control.
Seismometer feedforward filter is a similar filter but is used for cancellation of the seismic noise by feeding seismometer signal to actuators.
Feedback controllers are digital filters that convert sensing signals to actuation signals to achieve feedback control, which minimizes the displacement level of a controlled platform to achieve active isolation.
While not shown in this paper, we claim that all of these filter design problems can be treated as a complementary filter design problem because they are all optimization problems seeking an optimal trade-off between two frequency dependent quantities such as seismic noise and sensor noise.
So, they can all be solved using the method provided for synthesizing complementary filters.
These problems will be solved and demonstrated in future work.

Although it is shown that optimal complementary filters can be synthesized for a sensor fusion configuration with two sensors, active isolation systems can utilize even more sensors.
For example, some active isolation platforms in LIGO are equipped with relative displacement sensors, geophones, and seismometers~\cite{Matichard:2015eva}.
This requires a low-pass filter, a band-pass filter, and a high-pass filter for sensor fusion of these three sensors.
Although it was shown that $\mathcal{H}_\infty$ methods can be used for synthesizing complementary filters for any arbitrary number of sensors if frequency-dependent specifications are given~\cite{dehaeze19_compl_filter_shapin_using_synth}, the minimization of super sensor noise in a three-sensor configuration was not.
Therefore, it remains as a future work and will be studied as an extension of this paper.

	\section{Conclusion\label{sec:conclusion}}
Sensor fusion is a technique that combines multiple sensors into one super sensor that has better noise performance.
Complementary filters are used for sensor fusion in active isolation systems in gravitational-wave detectors.
While conventional designs of complementary filters can be sub-optimal and irreproducible, a method is proposed to synthesize complementary filters in the $\mathcal{H}_\infty$-optimal sense.
The generated complementary filters minimize the noise of the super sensor at all frequencies, making it equally close to the lower bound at all frequencies in the logarithmic sense.
The proposed method only uses information of the sensor noise and requires minimal human intervention.
The effectiveness of the synthesis was demonstrated in sensor fusion application using typical noises of sensors used in current gravitational wave detectors.
It was shown that the method gives complementary filters that perform better than a pre-designed one and necessary features in the filters, such as notches, can be generated naturally as a result of optimization.
Also, it was shown that pre-designed filters cannot be reused in a new environment while this method adapts and makes new complementary filters that are optimized for the new environment.
Other control problems in active isolation systems, such as sensor correction, feedforward, and feedback control problems, can be treated as complementary filter problems and hence can be solved using the same method.
With the $\mathcal{H}_\infty$ method, current GW detectors can be benefited from improved control performance and optimal control filters for active isolation can be designed easily for upcoming GW detectors such as the Einstein Telescope.

	\ack
The authors thank Fabi$\acute{\rm{a}}$n Erasmo Pe$\tilde{\mathrm{n}}$a Arellano, Lucia Trozzo, Katherine Dooley, Joris van Heijningen, and the Einstein Telescope Instrument Science Board for insightful suggestions and comments, which have greatly improved our work.
The work described in this paper is partially supported by grants The Croucher Foundation of Hong Kong.
	\section*{References}
	\bibliographystyle{unsrt}
	\bibliography{bibliography_revised}

\begin{thebibliography}{10}

\bibitem{aligo}
J.~Aasi et~al.
\newblock Advanced {LIGO}.
\newblock {\em Classical and Quantum Gravity}, 32(7):074001, mar 2015.

\bibitem{Acernese_2014}
F~Acernese et~al.
\newblock Advanced virgo: a second-generation interferometric gravitational
  wave detector.
\newblock {\em Classical and Quantum Gravity}, 32(2):024001, dec 2014.

\bibitem{10.1093/ptep/ptx180}
T~Akutsu et~al.
\newblock {Construction of KAGRA: an underground gravitational-wave
  observatory}.
\newblock {\em Progress of Theoretical and Experimental Physics}, 2018(1), 01
  2018.
\newblock 013F01.

\bibitem{Akutsu_2019}
T~Akutsu et~al.
\newblock First cryogenic test operation of underground km-scale
  gravitational-wave observatory {KAGRA}.
\newblock {\em Classical and Quantum Gravity}, 36(16):165008, jul 2019.

\bibitem{quadpaper}
N~Robertson, Cagnoli Gianpietro, D~Crooks, E~Elliffe, J.~Faller, P~Fritschel,
  S~Goßler, A~Grant, A~Heptonstall, James Hough, Harald Lück, R~Mittleman,
  Michael Perreur-Lloyd, M~Plissi, S~Rowan, David Shoemaker, P~Sneddon,
  K~Strain, C~Torrie, and P~Willems.
\newblock Quadruple suspension design for advanced ligo.
\newblock {\em Classical and Quantum Gravity}, 19:4043, 07 2002.

\bibitem{update_quadruple}
S~M Aston, M~A Barton, A~S Bell, N~Beveridge, B~Bland, A~J Brummitt, G~Cagnoli,
  C~A Cantley, L~Carbone, A~V Cumming, L~Cunningham, R~M Cutler, R~J~S
  Greenhalgh, G~D Hammond, K~Haughian, T~M Hayler, A~Heptonstall, J~Heefner,
  D~Hoyland, J~Hough, R~Jones, J~S Kissel, R~Kumar, N~A Lockerbie, D~Lodhia,
  I~W Martin, P~G Murray, J~O'Dell, M~V Plissi, S~Reid, J~Romie, N~A Robertson,
  S~Rowan, B~Shapiro, C~C Speake, K~A Strain, K~V Tokmakov, C~Torrie, A~A van
  Veggel, A~Vecchio, and I~Wilmut.
\newblock Update on quadruple suspension design for advanced {LIGO}.
\newblock 29(23):235004, oct 2012.

\bibitem{Braccini:2000hug}
S.~Braccini, C.~Casciano, V.~Dattilo, F.~Frasconi, A.~Gaddi, A.~Giazotto,
  G.~Losurdo, R.~Passaquieti, P.~Ruggi, and A.~Vicere.
\newblock {Performances Of The R\&D Super-attenuator (SA) Chain Of The VIRGO
  Experiment}.
\newblock In {\em {34th Rencontres de Moriond: Gravitational Waves and
  Experimental Gravity}}, pages 189--198, Hanoi, 2000. The Gioi world
  Publishers.

\bibitem{KAGRA:2019ywe}
Y.~Akiyama et~al.
\newblock {Vibration isolation system with a compact damping system for power
  recycling mirrors of KAGRA}.
\newblock {\em Class. Quant. Grav.}, 36(9):095015, 2019.

\bibitem{typeb_paper}
T~Akutsu et~al.
\newblock {Vibration isolation systems for the beam splitter and signal
  recycling mirrors of the {KAGRA} gravitational wave detector}.
\newblock 38(6):065011, mar 2021.

\bibitem{Matichard:2015eva}
F.~Matichard et~al.
\newblock {Seismic isolation of Advanced LIGO: Review of strategy,
  instrumentation and performance}.
\newblock {\em Class. Quant. Grav.}, 32(18):185003, 2015.

\bibitem{microseism}
Fabrice Ardhuin and Lucia Gualtieri.
\newblock {How ocean waves rock the Earth: Two mechanisms explain microseisms
  with periods 3 to 300 s}.
\newblock {\em Geophysical Research Letters}, 42, 01 2015.

\bibitem{lens.org/141-760-662-241-570}
Shinji Miyoki.
\newblock {Current status of KAGRA}.
\newblock In {\em Ground-based and Airborne Telescopes VIII}, volume 11445,
  pages 192--204. SPIE, December 2020.
\newblock
  \url{https://www.spiedigitallibrary.org/conference-proceedings-of-spie/11445/114450Z/Current-status-of-KAGRA/10.1117/12.2560824.full}
  ; \url{https://ui.adsabs.harvard.edu/abs/2020SPIE11445E..0ZM/abstract}.

\bibitem{LIGO:2020pzq}
E.~Schwartz et~al.
\newblock {Improving the robustness of the advanced LIGO detectors to
  earthquakes}.
\newblock {\em Class. Quant. Grav.}, 37(23):235007, 2020.

\bibitem{inclinometer}
N.~Azaryan, Julian Budagov, V.~Glagolev, M.~Lyablin, Andrei Pluzhnikov,
  A.~Seletsky, G.~Trubnikova, Beniamino Di~Girolamo, Jean-Christophe Gayde, and
  D.~Mergelkuhlb.
\newblock {Professional Precision Laser Inclinometer: the Noises Origin and
  Signal Processing}.
\newblock {\em Physics of Particles and Nuclei Letters}, 16:264--276, 05 2019.

\bibitem{compact_interferometer}
S~J Cooper, C~J Collins, A~C Green, D~Hoyland, C~C Speake, A~Freise, and C~M
  Mow-Lowry.
\newblock {A compact, large-range interferometer for precision measurement and
  inertial sensing}.
\newblock 35(9):095007, mar 2018.

\bibitem{2019_6d_interferometric_inertial_isolation_system}
C~M Mow-Lowry and D~Martynov.
\newblock {A 6D interferometric inertial isolation system}.
\newblock 36(24):245006, nov 2019.

\bibitem{robert12_introd_random_signal_applied_kalman}
Patrick Y. C.~Hwang Robert Grover~Brown.
\newblock {\em Introduction to Random Signals and Applied Kalman Filtering with
  Matlab Exercises}.
\newblock Wiley, 4 edition, 2012.

\bibitem{brown72_integ_navig_system_kalman_filter}
R.~G. Brown.
\newblock Integrated navigation systems and kalman filtering: a perspective.
\newblock {\em Navigation}, 19(4):355--362, 1972.

\bibitem{plummer06_optim_compl_filter_their_applic_motion_measur}
A.~R. Plummer.
\newblock Optimal complementary filters and their application in motion
  measurement.
\newblock {\em Proceedings of the Institution of Mechanical Engineers, Part I:
  Journal of Systems and Control Engineering}, 220(6):489--507, 2006.

\bibitem{hua05_low_ligo}
Wensheng Hua.
\newblock {\em Low frequency vibration isolation and alignment system for
  advanced LIGO}.
\newblock PhD thesis, stanford university, 2005.

\bibitem{Hua04polyphasefir}
Wensheng Hua, Dan~B. Debra, Corwin~T. Hardham, Brian~T. Lantz, and Joseph~A.
  Giaime.
\newblock {Polyphase FIR Complementary Filters for Control Systems}.
\newblock In {\em Proceedings of ASPE Spring Topical Meeting on Control of
  Precision Systems}, pages 109--114, 2004.

\bibitem{ligo_earthquake_mode}
Sebastien Biscans, Jim Warner, Richard Mittleman, Christopher Buchanan, Michael
  Coughlin, Matthew Evans, Hunter Gabbard, Jan Harms, Brian Lantz, Nikhil
  Mukund~Menon, Arnaud Pele, Charles Pezerat, Pascal Picart, Hugh Radkins, and
  Thomas Shaffer.
\newblock Control strategy to limit duty cycle impact of earthquakes on the
  ligo gravitational-wave detectors.
\newblock {\em Classical and Quantum Gravity}, 35, 07 2017.

\bibitem{Sekiguchi:2016cdw}
Takaori Sekiguchi.
\newblock {\em {A Study of Low Frequency Vibration Isolation System for Large
  Scale Gravitational Wave Detectors}}.
\newblock PhD thesis, Tokyo U., 2016.

\bibitem{Heijningen2018TurnUT}
Joris van Heijningen.
\newblock {\em Turn up the bass!: Low-frequency performance improvement of
  seismic attenuation systems and vibration sensors for next generation
  gravitational wave detectors}.
\newblock PhD thesis, 2018.

\bibitem{lucia_thesis}
Lucia Trozzo.
\newblock {\em Low Frequency Optimization and Performance of Advanced Virgo
  Seismic Isolation System}.
\newblock PhD thesis, 2018.

\bibitem{fujii_thesis}
Yoshinori Fujii.
\newblock {\em Fast localization of coalescing binaries with gravitational wave
  detectors and low frequency vibration isolation for KAGRA}.
\newblock PhD thesis, 2019.

\bibitem{ET_design_report_update_2020}
ET~Editorial Team.
\newblock Design report update 2020 for the einstein telescope.
\newblock ET Docs: ET-0007B-20. Available at
  \url{https://apps.et-gw.eu/tds/?content=3&r=17245}.

\bibitem{pso_sensor_correction}
Jonathan~J. {Carter}, Samuel~J. {Cooper}, Edward {Thrift}, Joseph {Briggs}, Jim
  {Warner}, Michael~P. {Ross}, and Conor~M. {Mow-Lowry}.
\newblock {Particle swarming of sensor correction filters}.
\newblock {\em Classical and Quantum Gravity}, 37(20):205009, October 2020.

\bibitem{dehaeze19_compl_filter_shapin_using_synth}
Thomas Dehaeze, Mohit Vermat, and Collette Christophe.
\newblock Complementary filters shaping using $\mathcal{H}_\infty$ synthesis.
\newblock In {\em 7th International Conference on Control, Mechatronics and
  Automation (ICCMA)}, pages 459--464, 2019.

\bibitem{zames_hinfinity}
G.~Zames.
\newblock {Feedback and optimal sensitivity: Model reference transformations,
  multiplicative seminorms, and approximate inverses}.
\newblock {\em IEEE Transactions on Automatic Control}, 26(2):301--320, 1981.

\bibitem{Glover1988StatespaceFF}
Keith Glover and John~C. Doyle.
\newblock {State-space formulae for all stabilizing controllers that satisfy
  and H$\infty$ norm bound and relations to risk sensitivity}.
\newblock {\em Systems \& Control Letters}, 11:167--172, 1988.

\bibitem{PACKARD1992271}
Andy Packard, Kemin Zhou, Pradeep Pandey, Jorn Leonhardson, and Gary Balas.
\newblock {Optimal, constant I/O similarity scaling for full-information and
  state-feedback control problems}.
\newblock {\em Systems \& Control Letters}, 19(4):271--280, 1992.

\bibitem{python_control}
{Python Control Systems Library}.
\newblock \url{http://python-control.org}.

\bibitem{Benner1999}
Peter Benner, Volker Mehrmann, Vasile Sima, Sabine Van~Huffel, and Andras
  Varga.
\newblock {\em SLICOT---A Subroutine Library in Systems and Control Theory},
  pages 499--539.
\newblock Birkh{\"a}user Boston, Boston, MA, 1999.

\bibitem{fortran77}
Petko Petkov, Da-Wei Gu, and M.M. Konstantinov.
\newblock {Fortran 77 Routines for H$\infty$ and H2 Design of Discrete-Time
  Linear Control Systems}.
\newblock 06 1999.

\bibitem{kontrol}
{Kontrol}.
\newblock \url{kontrol.readthedocs.io}.

\bibitem{skogestad}
Sigurd Skogestad and I~Postlethwaite.
\newblock {\em {Multivariable Feedback Control: Analysis and Design}},
  volume~2.
\newblock 01 2005.

\bibitem{nelder_mead}
J.~A. Nelder and R.~Mead.
\newblock {A Simplex Method for Function Minimization}.
\newblock {\em The Computer Journal}, 7(4):308--313, 01 1965.

\bibitem{powell}
M.~J.~D. Powell.
\newblock {An efficient method for finding the minimum of a function of several
  variables without calculating derivatives}.
\newblock {\em The Computer Journal}, 7(2):155--162, 01 1964.

\bibitem{differential_evolution}
R.~Storn and K.~Price.
\newblock {Differential Evolution – A Simple and Efficient Heuristic for
  global Optimization over Continuous Spaces}.
\newblock {\em Journal of Global Optimization}, 11:341--359, 12 1997.

\end{thebibliography}
	
\end{document}